\documentclass[longauth,twocolumn,traditabstract]{aa}
\usepackage{txfonts}
\usepackage[dvips]{graphics}
\usepackage[dvips]{graphicx}
\usepackage{graphicx}
\usepackage{longtable}
\usepackage{color}
\usepackage{url}
\usepackage{dcolumn} 

\usepackage{natbib}

\newcommand\Mrk{Mrk~421}
\newcommand\PKSb{PKS~2005-489}
\newcommand\PKS{PKS~2155-304}
\newcommand\ESn{1ES~0229+200}
\newcommand\Hm{H~2356-309}
\newcommand\ESu{1ES~1101-232}
\newcommand\ESs{1ES~0347-121}
\newcommand{\hess}{\textsc{H.E.S.S.}}
\newcommand{\dgr}{\ensuremath{^\mathrm{o}}}

\bibpunct{(}{)}{;}{a}{}{,} 

\begin{document}

\title{Measurement of the extragalactic background light imprint on the spectra of the brightest blazars observed with \hess}

\author{H.E.S.S. Collaboration
\and A.~Abramowski \inst{1}
\and F.~Acero \inst{2}
\and F.~Aharonian \inst{3,4,5}
\and A.G.~Akhperjanian \inst{6,5}
\and G.~Anton \inst{7}
\and S.~Balenderan \inst{8}
\and A.~Balzer \inst{7}
\and A.~Barnacka \inst{9,10}
\and Y.~Becherini \inst{11,12}
% ...
}

\institute{
Universit\"at Hamburg, Institut f\"ur Experimentalphysik, Luruper Chaussee 149, D 22761 Hamburg, Germany \and
Laboratoire Univers et Particules de Montpellier, Universit\'e Montpellier 2, CNRS/IN2P3,  CC 72, Place Eug\`ene Bataillon, F-34095 Montpellier Cedex 5, France \and
Max-Planck-Institut f\"ur Kernphysik, P.O. Box 103980, D 69029 Heidelberg, Germany \and
Dublin Institute for Advanced Studies, 31 Fitzwilliam Place, Dublin 2, Ireland \and
National Academy of Sciences of the Republic of Armenia, Yerevan  \and
Yerevan Physics Institute, 2 Alikhanian Brothers St., 375036 Yerevan, Armenia \and
Universit\"at Erlangen-N\"urnberg, Physikalisches Institut, Erwin-Rommel-Str. 1, D 91058 Erlangen, Germany \and
University of Durham, Department of Physics, South Road, Durham DH1 3LE, U.K. \and
Nicolaus Copernicus Astronomical Center, ul. Bartycka 18, 00-716 Warsaw, Poland \and
CEA Saclay, DSM/Irfu, F-91191 Gif-Sur-Yvette Cedex, France \and
APC, AstroParticule et Cosmologie, Universit\'{e} Paris Diderot, CNRS/IN2P3, CEA/Irfu, Observatoire de Paris, Sorbonne Paris Cit\'{e}, 10, rue Alice Domon et L\'{e}onie Duquet, 75205 Paris Cedex 13, France,  \and
Laboratoire Leprince-Ringuet, Ecole Polytechnique, CNRS/IN2P3, F-91128 Palaiseau, France
% ...
}

\authorrunning{\hess\ collaboration}
\titlerunning{The EBL imprint on H.E.S.S. spectra}
\date{Received 10/09/2012 / Accepted 04/12/2012}

\abstract{
The extragalactic background light (EBL) is the diffuse radiation with the second highest energy density in the Universe after the cosmic microwave background. The aim of this study is the measurement of the imprint of the EBL opacity to $\gamma$-rays on the spectra of the brightest extragalactic sources detected with the High Energy Stereoscopic System (H.E.S.S.). The originality of the method lies in the joint fit of the EBL optical depth and of the intrinsic spectra of the sources, assuming intrinsic smoothness. Analysis of a total of $\sim10^5$ $\gamma$-ray events enables the detection of an EBL signature at the 8.8$\sigma$ level and constitutes the first measurement of the EBL optical depth using very-high energy ($E>100$~GeV) $\gamma$-rays. The EBL flux density is constrained over almost two decades of wavelengths [0.30~$\mu$m,~17~$\mu$m] and the peak value at $1.4~\mu$m is derived as $\lambda\rm{F}_\lambda=15\pm2_{\rm stat}\pm3_{\rm sys}$~nW~m$^{-2}$~sr$^{-1}$. 
}

\offprints{\\Jonathan Biteau - email: biteau(at)in2p3.fr\\
Berrie Giebels - email: berrie(at)in2p3.fr\\
David Sanchez - email: david.sanchez(at)mpi-hd.mpg.de}
\keywords{Gamma rays: galaxies -- Cosmology: cosmic background radiation -- Galaxies: BL Lacertae objects: individual: Mrk~421, PKS~2005-489, PKS~2155-304, 1ES~0229+200, H~2356-309, 1ES~1101-232, 1ES~0347-121}
\maketitle

\section{Introduction}
\label{Intro}
The extragalactic background light (EBL) is the second most intense diffuse radiation in the Universe, and its spectral energy distribution is composed of two bumps: the cosmic optical background (COB) and the cosmic infra-red background (CIB). The former is mainly due to the radiation emitted by stellar nucleosynthesis in the optical (O) to near infrared (IR), while the latter stems from UV-optical light absorbed and re-radiated by dust in the IR domain \citep[for a review, see ][]{SummaryObs1}.

Direct measurements of the EBL flux density prove to be difficult, mainly because foreground contamination, e.g. by the zodiacal light, can result in an overestimation. Strict lower limits have been derived from integrated galaxy counts \citep[see, e.g.,][for more details]{2000MNRAS.312L...9M,2004ApJS..154...39F,SummaryObs2}. The limits derived from direct measurement in the near IR domain typically are one order of magnitude above the lower limits from source counts.

Strong constraints on the EBL density are derived using extragalactic very high energy (VHE, $E>100$~GeV) $\gamma$-ray sources. VHE $\gamma$-rays interact with O-IR photons via electron-positron pair production, resulting in an attenuated flux that is detected on Earth \citep{REF::NIKISHOV::JETP1962,1966PhRvL..16..479J,1967PhRv..155.1408G}. Assuming that there is no intrinsic break in the energy range of interest \citep[as in][]{1992ApJ...390L..49S} and that the hardness of the spectrum is limited, stringent upper limits on the EBL opacity to $\gamma$-rays have been derived \citep[e.g.][]{EBLAHA,EBLMaRa}. Studies exploiting Fermi-LAT measurements as templates of the intrinsic spectra have also recently been performed \citep{Georganopoulos,Orr,Meyer12}. Current models of the EBL are in close agreement with these limits, and they converge on a peak value of the stellar component $\lambda{\rm F}_\lambda~\sim~12$~nW~m$^{-2}$~sr$^{-1}$, yielding a consistent value for the opacity to $\gamma$-rays \citep[see, e.g.,][]{Dominguez}. 

The attenuation by the EBL is expected to leave a unique, redshift dependent and energy dependent imprint on the VHE spectra. While at energies above $E\gtrsim 5$ to 10~TeV (depending on the redshift of the source) a sharp cut-off is expected resulting from the CIB, a weaker modulation should imprint the spectra in the energy range between $\sim100$~GeV and $\sim5-10$~TeV, resulting from the rise and fall of the first peak of the EBL, the COB \citep{1999A&A...349...11A,2003A&A...403..523A}. A significant detection of this modulation, localized in a relatively narrow energy range, requires studying high-quality spectra, as, e.g., measured during the strong flux outburst of PKS 2155-304 in 2006 \citep{2007ApJ...664L..71A}, under the assumption that the intrinsic spectra are smooth over the energy range being studied.

This signature is searched for in the spectra of the brightest extragalactic blazars detected by \hess\ with a maximum likelihood method, leaving the parameters of the intrinsic spectra free. The originality and the strength of the technique lie in the joint fit of the EBL optical depth and of the intrinsic spectra of the sources, fully accounting for intrinsic curvature. This derivation of the EBL optical depth with \hess\ data does not rely on constraints on the intrinsic spectrum from assumptions about the acceleration mechanism and results in a {\em measurement} of the optical depth, compared to the upper limits derived in previous studies.

The sample of blazars studied in this paper, the data analysis, and the spectral fitting method are described in Sect.~\ref{DataAnalysis}. In Sect.~\ref{Results}, the results are presented and the systematic uncertainties are discussed. Finally, the results of this analysis are compared with the current constraints in Sect.~\ref{Discussion}.

\begin{table}
\centering
\begin{tabular}{l c c c }
\hline\hline
Data set & $N_\gamma$ & $\sigma$ & $E_{\rm min}-E_{\rm max}$ \\
 & & & [TeV]\\
\hline
\Mrk\ (\emph 1) & 3381 & 96.7 & $0.95-41$ \\
\Mrk\ (\emph 2) & 5548 & 135 & $0.95-37$ \\
\Mrk\ (\emph 3) & 5156 & 134 & $0.95-45$ \\
\PKSb\ (\emph 1) & 1540 & 25.3 & $0.16-37$  \\
\PKSb\ (\emph 2) & 910 & 28.9 & $0.18-25$  \\
\PKS\ (\emph{2008}) & 5279 & 99.2 & $0.13-19$ \\
\PKS\ (\emph 1) & 3499 & 93.0 & $0.13-5.7$  \\
\PKS\ (\emph 2) & 3470 & 116 & $0.13-9.3$  \\
\PKS\ (\emph 3) & 9555 & 186 & $0.13-14$  \\
\PKS\ (\emph 4) & 4606 & 132 & $0.18-4.6$  \\
\PKS\ (\emph 5) & 11901 & 219 & $0.13-5.7$  \\
\PKS\ (\emph 6) & 6494 & 166 & $0.15-5.7$   \\
\PKS\ (\emph 7) & 8253 & 191 & $0.20-7.6$  \\
\ESn & 670 & 12.6 & $0.29-25$   \\
\Hm & 1642 & 21.2 & $0.11-34$  \\
\ESu & 1268 & 17.8 & $0.12-23$ \\
\ESs & 604 & 13.5 & $0.13-11$  \\
\hline
\end{tabular}
\caption{Data sets on VHE blazars detected by \hess\ that are used for this study of the EBL. For highly variable sources, the data are divided into smaller subsets that are indexed in column 1 and correspond to restricted flux ranges. The photon excess, detection significance, and energy range of the spectra (in TeV) are given in columns 2, 3, and 4, respectively.}
\label{DetectedBlazars}
\end{table}

\section{Analysis of \hess\ data}\label{DataAnalysis}
\subsection{Reduction of \hess\ data}
The high energy stereoscopic system (\hess) is an array of four imaging atmospheric Cherenkov telescopes located 1800 m above sea level, in the Khomas Highland, Namibia (23\dgr16'18''S, 16\dgr30'01''E). The Cherenkov light emitted by VHE-particle-induced showers in the atmosphere is focussed with 13 m diameter optical reflectors onto ultra fast cameras \citep{Bernlohr,Hinton2004}. Each camera consists of 960 photomultipliers equipped with Winston cones to maximize the collection of light. The coincident detection of a shower with at least two telescopes improves the $\gamma$/hadron separation \citep{Funk,Crab}. 

The data sets studied in this paper were selected with standard quality criteria \citep[weather and stability of the instruments as in][]{Crab}, and the main analysis was performed with {\it Model analysis} \citep{MathieuANA}. Based on a maximum likelihood method that compares the recorded images with simulated $\gamma$-rays, this analysis improves the $\gamma$/hadron separation with respect to the standard Hillas analysis method \citep[see e.g.][]{Crab}, especially for low energies. 

The lowest photo-electron threshold of 40 p.e. per camera image after cleaning \citep[Loose Cuts,][]{MathieuANA} was adopted to cover the largest possible energy range. The on-events were taken from circular regions around the sources with a radius of 0.11\dgr. The background was estimated with the conventional reflected regions method \citep{Crab}. A minimum of three operating telescopes was required to derive the spectrum of a source, the redundancy allowing an improved reconstruction of the direction and energy of the $\gamma$-rays.

A cross-check was performed with the standard multi-variate analysis (MVA) described in \citet{TMVAHD} and an independent calibration, yielding consistent results.

\subsection{Sample of sources}
\label{sample}
The detection of a subtle absorption feature, such as the effect of the EBL, relies on spectra measured with great accuracy, motivating the study of the extensively observed, bright \hess\ blazars. A cut on the detection significance \citep{LiMa} of $10 \sigma$ yielded a sample of seven blazars: \Mrk, \PKSb, \PKS, \ESn, \Hm, \ESu, and \ESs. 

\Mrk\ is the first extragalactic source ever detected in the VHE energy domain \citep{Punch92}. This highly variable BL Lac object is observed by \hess\ at large zenith angles \citep{MrkHESS2004}, yielding a high energy threshold around 1~TeV but also, with a large effective area at higher energies, photons up to $\sim 40$~TeV. Thus, even considering the low redshift of the source, $z=0.031$ \citep{1975ApJ...198..261U}, the EBL significantly impacts its observed spectrum, with an optical depth $\tau(E=10~{\rm TeV},z=0.031)\sim 1$. 

\PKSb\ \citep[$z=0.071$,][]{1987ApJ...318L..39F} and \Hm\ \citep[$z=0.165$,][]{2009MNRAS.399..683J} are two blazars at the $\sim2\%$ of the Crab nebula flux level, detected by \hess\ since it went into operation \citep{2356,2005begins}. While the latter does not show any sign of spectral variability \citep{2010A&A...516A..56H}, an intensive observation campaign on the former revealed significant variations \citep{2005,2011A&A...533A.110H}. 

Together with \Hm, \ESu\ has already been used for EBL studies. With a measured photon index smaller than three for a redshift of 0.186 \citep{1989ApJ...345..140R}, the spectrum of this source largely contributed to the stringency of the upper-limit derived by \citet{EBLAHA}. A dedicated study published in 2007 did not reveal any significant flux variations over the observation period between 2004 and 2005 \citep{1101}. 

\PKS\ \citep[$z=0.116$,][]{1993ApJ...411L..63F} is the brightest extragalactic source in the Southern sky, and it has been widely studied with \hess\ \citep{2005A&A...430..865A,2005A&A...442..895A,2007ApJ...664L..71A,2155_2008,2009A&A...502..749A,2155_2010,2012arXiv1201.4135H}. It exhibited a spectacular flux outburst in July 2006 \citep{2007ApJ...664L..71A}, with a flux so high that the number of detected $\gamma$-rays exceeds by far the cumulated excess from all the other \hess\ extragalactic sources. This study focusses on the high statistics data set from July 2006 and from a multi-wavelength campaign performed in 2008, where the low state of the source was measured with high precision \citep{2155_2008}. These detailed high quality spectra of \PKS\ have not been used to set limits on the EBL so far and are responsible for the most stringent constraints derived in this paper.

\ESn\ and \ESs\ are characterized by their redshift of 0.14 \citep{1993ApJ...412..541S} and 0.188 \citep{2005ApJ...631..762W}, respectively. The spectra of these sources \citep{0229,0347}  confirmed the EBL limits set by \citet{EBLAHA}, and their light curves were compatible with constant flux. 

For each source, the redshift, excess, significance, and energy range of the detected $\gamma$-rays are shown in the first columns of Table~\ref{DetectedBlazars}. Blazars sometimes exhibit spectral changes correlated with the flux \citep[e.g.][]{2155_2010} that could result in a scatter of the absorption feature estimates. Spectral variations can be particularly important compared to statistical fluctuations for highly significant ($\gtrsim 30\sigma$) sources. To minimize this effect, the data from \PKS\ (high state), \Mrk, and \PKSb\ were divided into several bins in flux with roughly the same logarithmic width and a similar number of $\gamma$-rays, using data slices of 28 min duration (runs). This resulted in  7, 3, and 2 bins for the sources, respectively, which are ordered by increasing level of flux and are listed in brackets in Table~\ref{DetectedBlazars}. The observational conditions for the various data sets on a single source vary from one set to another, implying different energy ranges.

\subsection{Spectral analysis}
\label{SpecAna}
The spectral analysis of the data sets described in Sect.~\ref{sample} was performed taking the EBL absorption $e^{-\tau(E,z,n)}$ into account, where the optical depth depends on the EBL density $n$ and on the energy $E$ of the $\gamma$-rays, emitted by a source located at a redshift $z$. The EBL optical depth was scaled with a normalization factor $\alpha$, as in \cite{FermiEBL}, yielding a spectral model for each source:
\begin{equation}
\label{Eq:SpecMod}
\phi_z(E)=\phi^\alpha_{\rm int}(E)\times\exp(-\alpha\times\tau(E,z,n))
\label{SpectralForm}
\end{equation}
where $\phi^\alpha_{\rm int}(E)$ is the intrinsic spectrum of the source, i.e. the de-absorbed spectrum assuming an EBL optical depth scaled by $\alpha$. 

The template chosen for the EBL density $n$ is the model of \cite{Fr08}, hereafter FR08, which is representative of the current state of the art of EBL modelling and for which the optical depth is finely discretized in energy and redshift\footnote{The optical depth derived by FR08 is tabulated from $z_0=10^{-3}$ to $z_1=2$ in steps of $\delta z =10^{-3}$ and from $E_0=20$~GeV to $E_1=170$~TeV for 50 logarithmic steps. An interpolation in energy is performed for the spectral analysis.}. The EBL normalization factor $\alpha$, defined in Eq.~\ref{Eq:SpecMod}, is thus an estimator of the ratio between the measured and template opacities $\tau_{\rm measured}/\tau_{\rm FR08}$. The particular choice of optical depth modelling only has a minor impact on the reconstruction of the EBL flux density, and the systematic uncertainty resulting from this choice is estimated in Sect.~\ref{SysUn}.

The functional form of the intrinsic spectrum $\phi^\alpha_{\rm int}(E)$ assumed in this study is taken from very general considerations about the source physics. Blazars spectral energy distributions are indeed commonly described with a leptonic emission, e.g. with synchrotron self Compton models \citep{1985ApJ...298..128B}. In the VHE range, a smooth and concave spectrum is expected with the possible addition of a cut-off arising from the Klein-Nishina effect or a cut-off in the underlying electron distribution. The concurrent hadronic scenarios result in a smooth spectrum \citep[see e.g.][]{1993A&A...269...67M, 1999ApJ...510..188B, 2000NewA....5..377A} that closely resembles the leptonic spectra in the VHE range. At the first order, the intrinsic spectra are described with the most natural functional form for a non-thermal emission: a power law (PWL), i.e. a linear function in log-log scale. To test for the presence of intrinsic curvature, the next order of complexity is readily achieved using the log-parabola (LP), which is the equivalent of the parabola in log-log scale. The exponential cut-off hypothesis is also tested (EPWL), since expected on theoretical grounds, and since the order of the equivalent log-log polynomial would be too high, unreasonably widening the parameter space. The next order of complexity is simply achieved by generalizing the last two models, adding a cut-off to the LP (ELP), and smoothing ($\gamma<1$) or sharpening ($\gamma>1$) the cut-off of the EPWL (SEPWL). The exact choice of the intrinsic models, which are detailed in Table~\ref{table:model}, does not strongly affect the EBL measurement described hereafter, as shown in Appendix~\ref{appSys}.

In the following, deviations from concavity are assumed to arise from the EBL absorption term, which is a reasonable assumption as long as the scenarios concurrent to the leptonic emission do not mimic the energy and redshift dependence of the EBL optical depth \citep[but see also][for other probes such as flat spectrum radio quasars]{Reimer2007}. The energy dependence of the EBL absorption deviates from mere concavity \citep[e.g.][]{2010APh....34..245R}, and inflection points in the observed spectra, which depend on the redshift of the source, constitute the key imprint that is reconstructed in this study.

\begin{table}[h]
\centering
	\begin{tabular}{ l c l }
		\hline\hline
		Name & Abbrev. & Function\footnotemark \\
		\hline
		Power law & PWL & $\phi_0 (E/E_0)^{-\Gamma}$ \\
		Log parabola & LP & $\phi_0 (E/E_0)^{-a - b \log(E/E_0)}$ \\
		Exponential cut- & EPWL & $\phi_0 (E/E_0)^{-\Gamma} \exp(-E/E_{\rm cut})$ \\
		off power law &  &  \\
		Exponential cut- & ELP & $\phi_0 (E/E_0)^{-a - b \log(E/E_0)} \exp(-E/E_{\rm cut}) $ \\
		off log parabola &  &  \\
		Super exponential & SEPWL & $\phi_0 (E/E_0)^{-\Gamma} \exp\left(-(E/E_{\rm cut})^{\gamma}\right) $ \\
		cut-off power law &  &  \\
		\hline 
	\end{tabular}
\caption{Smooth functions describing the intrinsic spectra of the sources studied in this paper.}
\label{table:model}
\end{table}
\footnotetext{The reference energy $E_0$ is set to the decorrelation energy of the spectrum.}

To quantify the amplitude of the EBL signature on \hess\ spectra, the maximum likelihood method developed by \citet{FermiEBL} was adapted. Likelihood profiles were computed as a function of $\alpha$ for each data set and each smooth intrinsic spectral model given in Table~\ref{table:model}, with $\alpha$ ranging from 0 to 2.5. For each value of $\alpha$, the models $\phi_z(E)$ were fitted to the data with the intrinsic spectral parameters free in the minimization procedure.  The best fit maximum likelihood $\mathcal{L}$ was converted\footnote{The log-likelihood is set to zero if the measured number of events matches the expected number of events in each bin.} into an equivalent $\chi^2=-2 \log \mathcal{L}$ allowing the goodness of the fit to assessed with the conventional $\chi^2$ probability as a function of $\alpha$.

An unconventional procedure was set up to select a model. { It ensures that the intrinsic curvature is fully taken into account} and that the extrinsic curvature due to the EBL absorption is not overestimated. Normally, the model with fewest parameters would be used unless a model with one extra parameter is statistically preferred. Here, the model with the highest $\chi^2$ probability was selected, regardless of the value of $\alpha$ for which this maximum is reached. Cases where two models had comparable maximum $\chi^2$ probabilities are discussed in the following.

\begin{figure}[h]
\hspace{0.05cm}\includegraphics[width=0.495\linewidth]{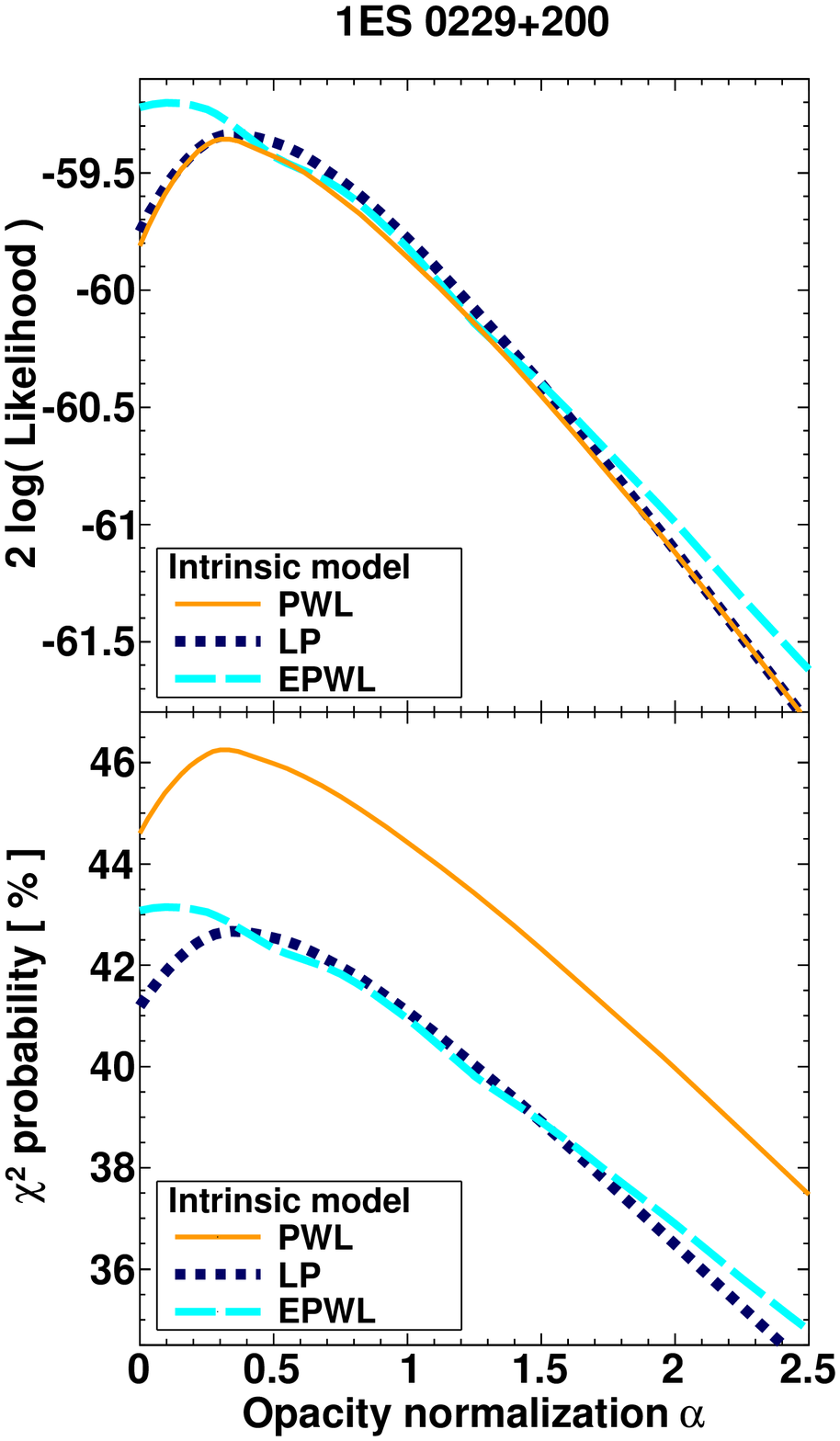}
\hspace{-0.07cm}\includegraphics[width=0.495\linewidth]{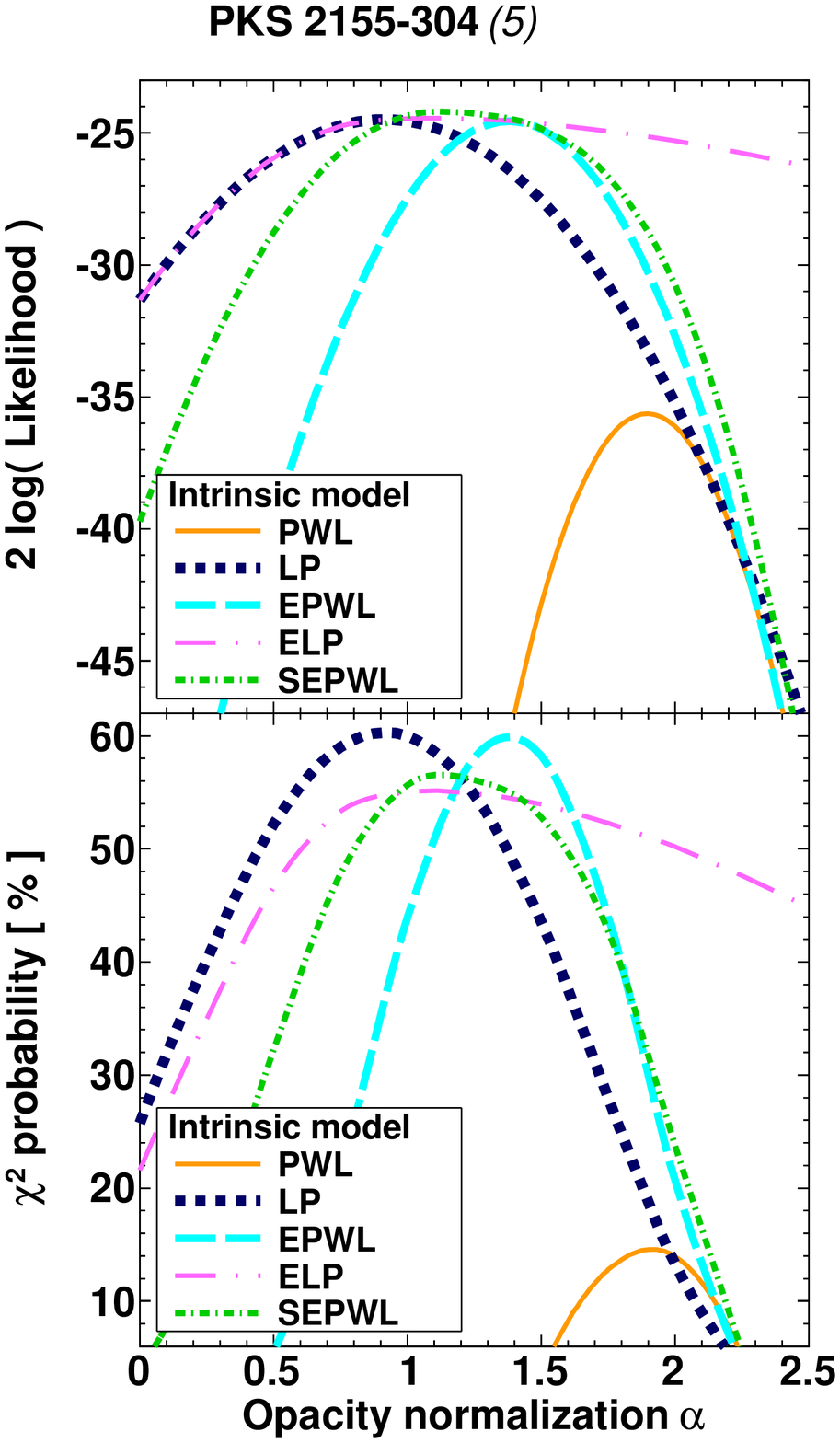}
\caption{{\it Top panels: } Likelihood profiles, as a function of the optical depth normalization for the different intrinsic models detailed in the legend. The examples of the data sets on \ESn\ and \PKS\ {\it (5)} are shown on the left and right, respectively.  {\it Bottom panels: } Corresponding $\chi^2$ probabilities as a function of the optical depth normalization. The PWL and SEPWL models are the spectral models chosen to describe the spectra of \ESn\ and  the fifth data set on \PKS, respectively.}
\label{fig:ExamplesProfile}
\end{figure}

Figure~\ref{fig:ExamplesProfile} shows the likelihood profiles and the $\chi^2$ probability profiles derived with the smooth intrinsic spectral models given in Table~\ref{table:model} for the data sets on \ESn\ and \PKS~{\it (5)}. In the first case, the likelihood profile derived with the PWL (two parameters) model does not significantly differ from those obtained with a LP (three parameters) or an EPWL (three parameters), but the decrease in the number of degrees of freedom with increasing complexity favours the PWL in term of $\chi^2$ probability, as shown in the bottom panel. In this case, the conventional method and our approach select the same model.

In the second case, corresponding to the fifth data set on \PKS, the LP and the EPWL significantly improve the fit compared to the PWL, in terms of maximum likelihood and of maximum $\chi^2$ probability. The LP profile and the EPWL profile have a similar maximum likelihood and, equivalently, (since the two models have the same number of free parameters) a similar maximum $\chi^2$ probability. Instead of performing an arbitrary choice between the LP and the EPWL, the profiles of the SEPWL, which generalizes the EPWL, and of the ELP, which generalize thes LP, were computed. According to our approach, the model with the highest maximum $\chi^2$ probability, in this case the SEPWL, was then selected.

As shown in Appendix~\ref{IntrinsicModelChoice}, using the common criteria with a significance level of 2$\sigma$ yields results in agreement with the unconventional method described here. The drawback of our approach is a less significant measurement with regards to the conventional method, due to the widening of the studied parameter space and the consequently larger statistical uncertainties. The intrinsic spectral models that were selected for each data set are given in Table~\ref{DetectedBlazarsSpec}. The impact of the selection of the intrinsic model on the final result is investigated in Appendix~\ref{IntrinsicModelChoice} and is included in the systematics (Sect.~\ref{SysUn}).

\begin{table}
\centering
\begin{tabular}{l c c c }
\hline\hline
Data set  & Spectral model & $\chi^2(\alpha_0)$ / $dof$ \\
\hline
\Mrk\ (\emph 1)  &  ELP & 21.5 / 31 \\
\Mrk\ (\emph 2)  &  ELP & 46.8 / 30 \\
\Mrk\ (\emph 3)  &  ELP & 34.8 / 28 \\
\PKSb\ (\emph 1) &  LP & 49.5 / 60 \\
\PKSb\ (\emph 2) &  LP & 31.8 / 46 \\
\PKS\ (\emph{2008})  & ELP & 21.9 / 37 \\
\PKS\ (\emph 1)  & PWL & 32.3 / 31 \\
\PKS\ (\emph 2)  & SEPWL & 25.3 / 28 \\
\PKS\ (\emph 3)  & SEPWL & 35.2 / 31 \\
\PKS\ (\emph 4)  & SEPWL & 19.1 / 21 \\
\PKS\ (\emph 5)  & SEPWL & 24.3 / 27 \\
\PKS\ (\emph 6)  & LP & 29.2 / 21 \\
\PKS\ (\emph 7)  & SEPWL & 13.6 / 13 \\
\ESn   & PWL & 60.1 / 60 \\
\Hm    & LP & 70.2 / 61\\
\ESu  & PWL & 62.6 / 69 \\
\ESs   & ELP & 31.7 / 35 \\
\hline
\end{tabular}
\caption{Spectral modelling of the data sets used to derive the likelihood profiles. The spectral models (see Sect.~\ref{SpecAna}) are given in column 2, where the acronyms PWL, LP, EPWL, ELP, and SEPWL are explained in Table~\ref{table:model}. The $\chi^2$ for the best fit EBL optical depth normalization $\alpha_0$ and the number of degrees of freedom $dof$ are given in column 3.}
\label{DetectedBlazarsSpec}
\end{table}

\section{Results}\label{Results}

\subsection{Measurement of the EBL optical depth normalization}

\begin{figure}[h]
\centering
\includegraphics[width=\linewidth]{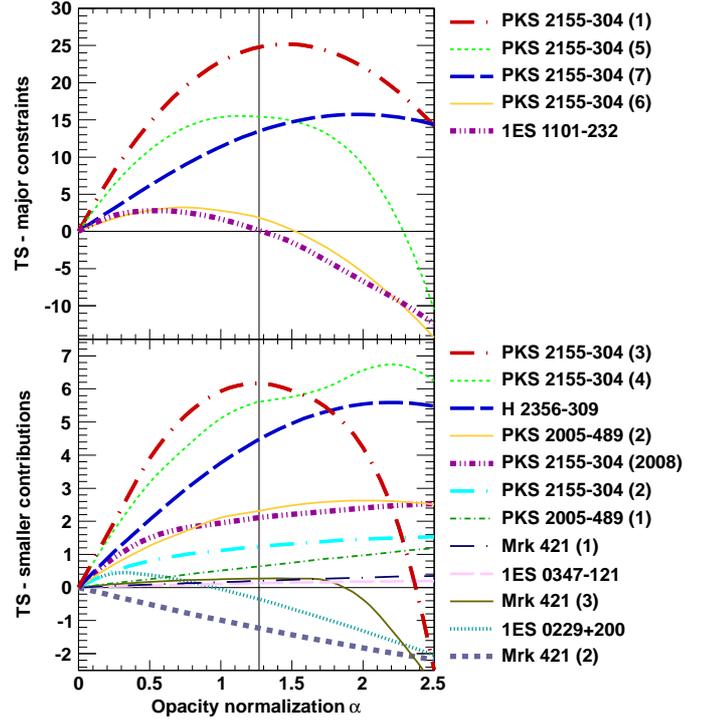}
\caption{Test statistic as a function of the normalized EBL optical depth for the intrinsic spectral models described in Table~\ref{DetectedBlazarsSpec}. The TS profiles are sorted by contribution to the measurement and the top panel shows the most constraining data sets, while the bottom panel shows the less constraining contributions. The vertical line indicates the best fit value derived in this study. Note the different scale on the vertical axis in the upper and the lower panel.}
\label{fig:IndivLikeli}
\end{figure}

For each data set, the likelihood $\mathcal{L}(\alpha)$ of the EBL optical depth normalization and of the intrinsic spectral model is compared to the hypothesis of a null EBL absorption $\mathcal{L}(\alpha=0)$. The test statistic (TS), defined by the likelihood ratio test ${\rm TS}=2\log\left[\mathcal{L}(\alpha=\alpha_0)/\mathcal{L}(\alpha=0)\right]$, is shown for each data set in Fig.~\ref{fig:IndivLikeli}.

With a total $\gamma$-ray excess of the order of 50,000 events, \PKS\ makes a major contribution to the EBL measurement. A maximum TS superior to 16 is achieved for the data sets \emph{(1)}, \emph{(5)}, and \emph{(7)}, meaning that a null EBL optical depth is rejected at the $4\sigma$ level by each of these data sets. An EBL optical depth scaled up by a factor two is rejected at the $3\sigma$ level by both the data set \emph{(6)} on \PKS\ and the one on \ESu. This constraint is not surprising since \ESu\ was already the most constraining source used by \citet{EBLAHA} to derive an upper-limit on the EBL opacity. The bottom panel of Fig.\ref{fig:IndivLikeli} shows the less constraining contributions. Though less significant individually, their combination contributes to roughly a third of the total TS and enables a null EBL optical depth to be rejected at the $\sim5\sigma$ level. 

\begin{figure}
\centering
\includegraphics[width=0.84\linewidth]{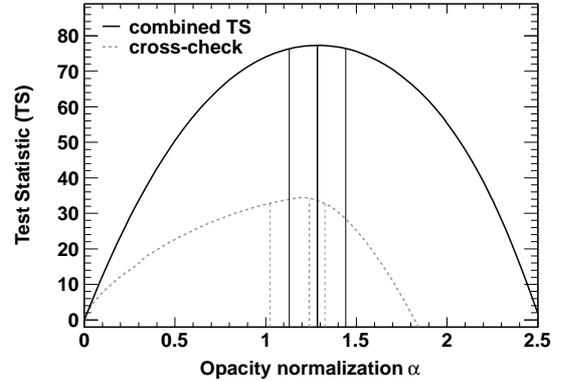}
\caption{Combined test statistic as a function of the normalized EBL optical depth. The results obtained with the {\it Model analysis} are shown with a black line and the cross-check led with the MVA analysis is shown with the dashed grey line. The best fit value and $1\sigma$ statistical uncertainties are shown with the vertical lines.}
\label{fig:SUMlikeli}
\end{figure}

The total TS  shown in Fig.~\ref{fig:SUMlikeli}, i.e. the sum of the individual ones presented in Fig.\ref{fig:IndivLikeli}, is maximum for $\alpha_0=1.27_{-0.15}^{+0.18}$, at $\sqrt{\Delta{\rm TS}} \sim 1.8\sigma$ above the unscaled FR08 template. The upper and lower standard deviations correspond to a variation of $\Delta{\rm TS}=1$ from the maximum test statistic TS~$=77.3$. The EBL optical depth template scaled up by a factor $\alpha_0$ is preferred at the $8.8\sigma$ level to a null optical depth, where the Gaussian significance is approximated by the square root of the likelihood ratio test. 

No outlier is present in the set of individual profiles, with best fit values of $1.44\pm0.29$ ($0.6\sigma$), $1.23\pm0.34$ ($0.1\sigma$), $1.97\pm0.48$ ($1.5\sigma$), $0.75\pm0.42$ ($1.2\sigma$), and $0.48\pm0.29$ ($1.6\sigma$) for the five most constraining data sets \PKS\ ({\it 1, 5, 7, 6}) and \ESu, respectively, where the number in brackets indicate the deviation to the best fit value $\alpha_0=1.27$. Similarly, the less constraining contributions do not differ by more than $\sqrt{{\rm TS_{max}-TS}(\alpha_0)} \lesssim1.5\sigma$ from the best fit value.

The total TS derived with the MVA analysis and an independent calibration is shown in Fig.~\ref{fig:SUMlikeli}. Though less significant, with a maximum TS of $33.9$, the best fit value $1.24_{-0.22}^{+0.09}$ is in close agreement with the optical depth normalization derived with the {\it Model analysis}. The larger energy range covered by the latter ($60\%$ wider in logarithmic scale) accounts for the difference in maximum TS.

\subsection{Redshift dependence}

To investigate the redshift dependence of the EBL optical depth normalization, the data set is divided by redshift in three groups. For \Mrk\ and \PKSb, the TS is maximum at $\alpha(z_1)=1.6^{+0.5}_{-1.1}$, for an average redshift of $z_1=0.051$. The TS of the eight data sets on \PKS\ ($z_2=0.116$) peaks at $\alpha(z_2)=1.36\pm0.17$. With the four other data sets, a maximum TS is obtained for $\alpha(z_3)=0.71^{+0.46}_{-0.29}$, corresponding to a mean redshift $z_3=0.170$.

Fitting the decreasing trend of the EBL normalization as a linear function of the redshift yields $\chi^2 / dof=0.41 / 1$, which does not significantly improve the fit with regards to a constant fit $\chi^2 / dof=1.83 / 2$. A likelihood ratio test prefers the linear fit only at the $1.1\sigma$ level. Any redshift dependence of the EBL normalization in the redshift range probed is therefore neglected in the following sections. 

\begin{figure}[h]
\includegraphics[width=0.95\linewidth]{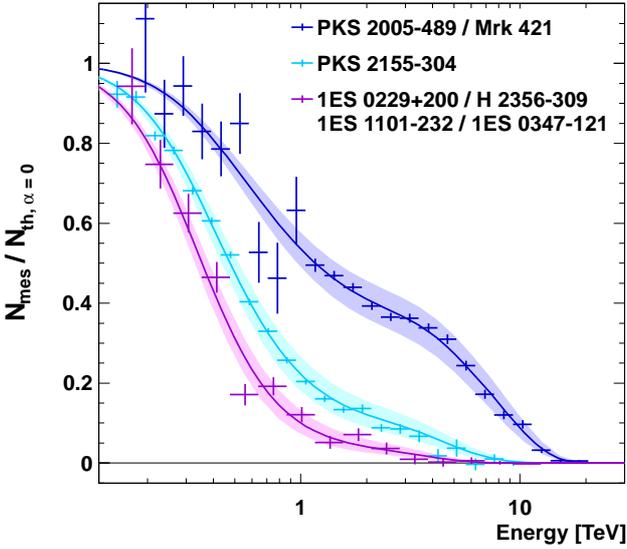}
\caption{Observed number of $\gamma$-rays over number of events expected from the intrinsic spectra vs $\gamma$-ray energy. The data sets are grouped by similar redshift and the detected and expected numbers of $\gamma$-rays are summed in each energy bin. The best fit EBL absorption is represented by the solid lines for the three redshifts corresponding to the groups of data sets and the shaded areas correspond to the $\pm1\sigma$ best fit EBL normalization.}
\label{fig:Res}
\end{figure}

Given { the limited amount of data}, the deviations from the best fit EBL normalization $\alpha_0=1.27_{-0.15}^{+0.18}$ can hardly be investigated at the single data set level. For the three above-mentioned groups of sources, the total number of measured events in each energy bin ($N_{\rm mes}$) is scaled to the expected number of events from the intrinsic spectra ($N_{\rm th,\ \alpha=0}$). This ratio is compared in Fig.~\ref{fig:Res} to the best fit model for the three average redshifts of 0.051, 0.116, and 0.17. Abrupt changes in the amplitude of the statistical uncertainties (e.g. around 1~TeV for the low redshift group: Mrk 421 / PKS 2005-489) are inherent to the grouping of data sets that cover different energy ranges (e.g. the data sets on Mrk 421 start at $\sim1$~TeV).

\subsection{Systematic uncertainty}
\label{SysUn}
An extensive investigation was undertaken of the systematic uncertainties arising from the method. Four sources of systematic uncertainties on the EBL optical depth normalization were identified: the analysis chain (background rejection, spectral analysis), the choice of intrinsic models and of the EBL template, as well as the limited knowledge of the energy scale due to the atmosphere. These systematic uncertainties are summarized in Table~\ref{table:sys} and detailed in Appendix~\ref{appSys}.

\begin{table}[h]
\centering
	\begin{tabular}{ c c }
		\hline\hline
		Sources of systematics & Estimated systematics \\
		\hline
		Analysis chain & 0.21 \\
		Intrinsic model & 0.10 \\
		EBL model & 0.06 \\
		Energy scale & 0.05 \\
		\hline
		Total & 0.25\\
		\hline 
	\end{tabular}
\caption{Sources of systematics and estimated uncertainties on the normalized EBL optical depth $\alpha_0=1.27_{ -0.15 \ \rm stat}^{ +0.18}$. A full discussion of the systematic uncertainties can be found in Appendix~\ref{appSys}.}
\label{table:sys}
\end{table}

The total systematic is estimated as $\sigma_{\rm sys}(\alpha_0)=0.25$ and is comparable to the statistical uncertainty on the normalized EBL optical depth $\alpha_0=1.27_{ -0.15 \ \rm stat}^{ +0.18}$.

\section{Discussion}
\label{Discussion}
The measurement of the EBL optical depth can be converted to an EBL flux density, but particular attention must be paid to the wavelength range covered. 

{ A $\gamma$-ray of energy $E^*$  and an EBL photon of energy $\epsilon^*$ tend to produce an electron-positron pair mostly for $E^*\epsilon^* =(2m_ec^2)^2$ \citep[peak of the cross section, see, e.g.,][]{1976tper.book.....J}}. The interaction can occur anywhere along the path of the $\gamma$-ray from the source and the relation for the EBL wavelength becomes, in the observer frame, 
\begin{equation}
\label{Eq:EBLWL}
( \lambda_{\rm EBL} / 1\ \mu{\rm m} )=1.187\times ( E / 1\ {\rm TeV} ) \times(1+z')^2
\end{equation}
with $z'<z$, where z is the redshift of the source and where $E$ is the $\gamma$-ray energy in the observer frame. To derive this relation between the EBL wavelength and the $\gamma$-ray energy, the width of the pair-creation cross-section as a function of energy is neglected. Taking it into account would result in an even wider wavelength coverage for a given $\gamma$-ray energy range.

\begin{figure}[b]
	\centering
	\includegraphics[width=1.0\linewidth]{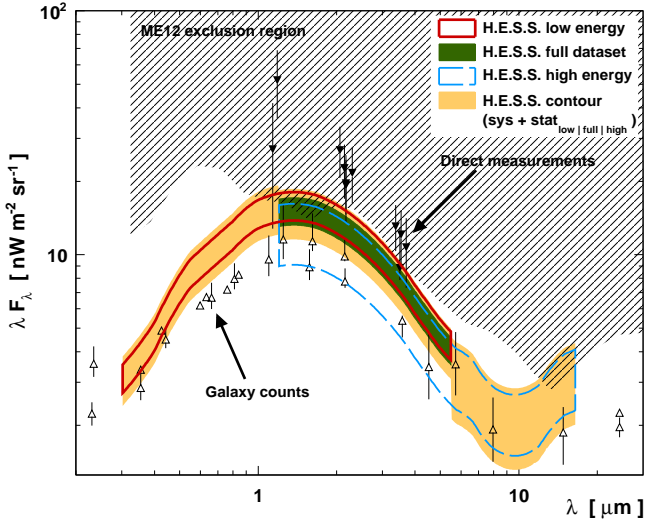}
	\caption{Flux density of the extragalactic background light versus wavelength. The 1$\sigma$ (statistical) contours derived for several energy ranges are described in the top-right legend. The systematic uncertainty is added quadratically to the statistical one to derive the \hess\ contour. Lower limits based on galaxy counts and direct measurements are respectively shown with empty upward and filled downward pointing triangles \citep[extracted from][]{Gilmore}. The region excluded by \cite{Meyer12} with VHE spectra is represented by the dashed area.}
	\label{fig:SEDebl}
\end{figure}

The detection of an EBL flux density scaled up by a factor $\alpha_0= 1.27_{ -0.15 \ \rm stat}^{ +0.18}\pm0.25_{\rm sys}$ is then valid in the overlap of the data-set energy ranges $[(1+z)^2E_{\rm min},E_{\rm max}]$, where the factor $(1+z)^2$ accounts for the redshift dependency in Eq.(\ref{Eq:EBLWL}). The measurement that is derived with all data sets is shown by the filled area in Fig.~\ref{fig:SEDebl} in the wavelength range $[1.2,5.5]~\mu$m, where $1.2~\mu$m (resp. $5.5~\mu$m) is the counterpart of the low (resp. high) energy bound of the \Mrk\ (resp. \PKS) data sets, as shown in Table~\ref{DetectedBlazarsWv}.

\begin{table}
\centering
\begin{tabular}{l c c c }
\hline\hline
Data set & $z$ & $E_{\rm min}-E_{\rm max}$  & $\lambda_{\rm min}-\lambda_{\rm max}$ \\
 &  &  [TeV] &  [$\mu$m] \\
\hline
\Mrk\ (\emph 1) & 0.031 & $0.95-41$ & $1.2 -49$ \\
\Mrk\ (\emph 2) & 0.031  & $0.95-37$ & $1.2 - 44$ \\ 
\Mrk\ (\emph 3) & 0.031  & $0.95-45$ & $1.2-53$   \\
\PKSb\ (\emph 1) & 0.071 & $0.16-37$ & $ 0.22-44$  \\
\PKSb\ (\emph 2) & 0.071 & $0.18-25$ & $0.25-30$  \\
\PKS\ (\emph{2008}) & 0.116 & $0.13-19$  & $0.30-23$ \\
\PKS\ (\emph 1) & 0.116 & $0.13-5.7$ & $0.19-6.8$ \\
\PKS\ (\emph 2) & 0.116 & $0.13-9.3$ & $0.19-11$  \\
\PKS\ (\emph 3) & 0.116 & $0.13-14$ & $0.19-17$  \\
\PKS\ (\emph 4) & 0.116 & $0.18-4.6$  & $0.19-5.5$ \\
\PKS\ (\emph 5) & 0.116 & $0.13-5.7$ & $0.27-6.8$ \\
\PKS\ (\emph 6) & 0.116 & $0.15-5.7$  & $0.19-6.8$ \\
\PKS\ (\emph 7) & 0.116 & $0.20-7.6$  & $0.22-9.0$ \\
\ESn & 0.14  & $0.29-25$  & $0.45-30$  \\
\Hm & 0.165  & $0.11-34$  & $0.18-40$ \\
\ESu & 0.186 & $0.12-23$ & $0.20-27$ \\
\ESs & 0.188 & $0.13-11$  & $0.22-13$ \\
\hline
\end{tabular}
\caption{EBL wavelength range probed by the data sets used in this study. The redshifts of the sources are given in column 2. The energy range of the spectra (in TeV) is given in column 3, and the EBL wavelengths probed with the subsets are given in column 4, where only the peak of the pair-creation cross-section is taken into account.}
\label{DetectedBlazarsWv}
\end{table}

To probe a wider wavelength range and to ensure the consistency of the modelling below and above $\sim 1~\mu$m, the TSs of data sets with comparable energy ranges were combined. Low EBL-wavelengths between 0.30 and $5.5~\mu$m were studied with the combination of the \ESs\ data set and the six \PKS\ data sets (\emph{1, 2, 4, 5, 6, 7}) while { the large EBL-wavelengths between 1.2 and $17~\mu$m were probed} by the \ESu, \ESn, \PKSb, \Mrk, \Hm\ data sets, and the two \PKS\ data sets (\emph{3, 2008}), all described in Table~\ref{DetectedBlazarsWv}. The normalized EBL optical depth measured in the various wavelength ranges and the corresponding EBL flux density are given in Table~\ref{tab:SED}.

\begin{table}[h]
\centering
\begin{tabular}{ c c c}
	\hline\hline
	$\tau_{\rm measured}/\tau_{\rm FR08}$  & $\lambda_{\rm min}$ \quad  -- \quad  $\lambda_{\rm max}$ & $\lambda {\rm F}_\lambda(\lambda_{\rm min})$ \quad  -- \quad  $\lambda {\rm F}_\lambda(\lambda_{\rm max})$\\
	 & $\mu$m & nW m$^{-2}$ sr$^{-1}$ \\
	\hline
	\\
	$1.27_{-0.15}^{+0.18}$ &1.2 \quad -- \quad  5.5 &  $14.8^{+2.1}_{-1.7}$\quad  -- \quad $4.0^{+0.6}_{-0.5}$ \\
	\\
	\hline
	\\
	$1.34_{-0.17}^{+0.19}$ & 0.30 \quad  --\quad  5.5 & $3.1 \pm 0.4$ \quad --\quad  $4.2^{+0.6}_{-0.5}$ \\
	\\
	$1.05_{-0.28}^{+0.32}$ & 1.2 \quad --\quad  17 & $12.2^{+3.7}_{-3.3}$ \quad --\quad  $3.2^{+1.0}_{-0.8}$ \\
	\\
	\hline 
\end{tabular}
	\caption{Measured normalization of the EBL optical depth, corresponding to the 1$\sigma$ (statistical) contours shown in Fig.~\ref{fig:SEDebl}. The second column indicates the wavelength range where this measurement is valid and the third column the corresponding flux densities. The first line corresponds to the full data set. The second and third lines indicate the value derived with smaller data sets focussed on specific energy ranges. The systematic uncertainty on the measurements listed in the first column is 0.25.}
\label{tab:SED}
\end{table}

The $1\sigma$ (statistical) contours of the EBL flux density for these two wavelength ranges and for the combination are compared in Fig.~\ref{fig:SEDebl} to other measurements and limits. The first peak of the EBL flux density, the COB, is entirely constrained by the low and the high energy data sets. The systematic uncertainty is quadratically added to the statistical uncertainty on the measurement with the full data set in the intermediate wavelength range, and to uncertainties on the low and high energy measurements in the extended ranges. The statistical uncertainties remain dominant around 10~$\mu$m.  In the UV to NIR domain, the systematic uncertainties, which are propagated from the optical depth normalization to the flux density as a single normalization factor, make a non-negligible contribution to the width of the contour. The detailed study of the dependence of the systematic uncertainties on the wavelength, based e.g. on deviations from the EBL template model, is beyond the scope of this paper but the comparison of various modellings in a complementary redshift band and wavelength range by \citet{2012arXiv1211.1671T} supports our choice of template.

The contours lie in between the constraints derived with galaxy counts and the direct measurements. A good agreement with the VHE upper limit derived by \citet{Meyer12} is also found over the wavelength range covered, with a maximum discrepancy between 1 and 2~$\mu$m smaller than the $1\sigma$ level. The analysis performed enables a measurement of the COB peak flux density of $\lambda\rm{F}_\lambda=15.0^{+2.1}_{-1.8}~\pm~2.8_{\rm sys}$~nW~m$^{-2}$~sr$^{-1}$ at $1.4\ \mu$m, where the peak value and uncertainties are derived by scaling up the FR08 EBL flux density by a factor $\alpha_0$. This value is compatible with the previous constraints on the EBL flux density derived with \hess\ data by \citet{EBLAHA}.

\section{Summary and conclusion}
The spectra of the brightest blazars detected by \hess\ were investigated for an EBL absorption signature. Assuming intrinsic spectral smoothness, the intrinsic spectral curvature was carefully disentangled from the EBL absorption effect. The EBL imprint is detected at an 8.8$\sigma$ level, which constitutes the first measurement of the EBL optical depth using VHE  $\gamma$-rays. The EBL flux density has been evaluated over almost two decades of wavelengths with a peak amplitude at $1.4\ \mu$m of $\lambda\rm{F}_\lambda=15 \pm 2_{\rm sys} \pm3_{\rm sys}$~nW~m$^{-2}$~sr$^{-1}$, in between direct measurements and lower limits derived with galaxy counts. 

The low energy threshold achieved with the upgrade of the \hess\ array, \hess~II, will enable the observation of the unabsorbed population of $\gamma$-rays and improve the constraints on the intrinsic spectra and thus on the absorption feature. The trough between the COB and the CIB will be characterized by the Cherenkov Telescope Array \citep[CTA,][]{2011ExA....32..193A} which will probe energies above 50~TeV. Finally, the increasing size of the sample of blazars detected at very high energies will improve the constraints on the redshift dependence of the EBL and establish a firm observational probe of the thermal history of the Universe.

\acknowledgements
The support of the Namibian authorities and of the University of Namibia
in facilitating the construction and operation of H.E.S.S. is gratefully
acknowledged, as is the support by the German Ministry for Education and
Research (BMBF), the Max Planck Society, the German Research Foundation (DFG), 
the French Ministry for Research,
the CNRS-IN2P3 and the Astroparticle Interdisciplinary Programme of the
CNRS, the U.K. Science and Technology Facilities Council (STFC),
the IPNP of the Charles University, the Czech Science Foundation, the Polish 
Ministry of Science and  Higher Education, the South African Department of
Science and Technology and National Research Foundation, and by the
University of Namibia. We appreciate the excellent work of the technical
support staff in Berlin, Durham, Hamburg, Heidelberg, Palaiseau, Paris,
Saclay, and in Namibia in the construction and operation of the
equipment.

\bibliography{EBL1.bbl}

\begin{thebibliography}{64}
\expandafter\ifx\csname natexlab\endcsname\relax\def\natexlab#1{#1}\fi

\bibitem[{{Abdo} {et~al.}(2010){Abdo}, {Ackermann}, {Ajello}, {Allafort},
  {Atwood}, {Baldini}, {Ballet}, {Barbiellini}, {Baring}, {Bastieri},
  {Baughman}, {Bechtol}, {Bellazzini}, {Berenji}, {Bhat}, {Blandford}, {Bloom},
  {Bonamente}, {Borgland}, {Bouvier}, {Brandt}, {Bregeon}, {Brez}, {Briggs},
  {Brigida}, {Bruel}, {Buehler}, {Burnett}, {Buson}, {Caliandro}, {Cameron},
  {Caraveo}, {Carrigan}, {Casandjian}, {Cavazzuti}, {Cecchi}, {{\c C}elik},
  {Charles}, {Chekhtman}, {Chen}, {Cheung}, {Chiang}, {Ciprini}, {Claus},
  {Cohen-Tanugi}, {Connaughton}, {Conrad}, {Costamante}, {Dermer}, {de
  Angelis}, {de Palma}, {Digel}, {Dingus}, {Silva}, {Drell}, {Dubois},
  {Favuzzi}, {Fegan}, {Finke}, {Fortin}, {Fukazawa}, {Funk}, {Fusco},
  {Gargano}, {Gasparrini}, {Gehrels}, {Germani}, {Giglietto}, {Gilmore},
  {Giommi}, {Giordano}, {Giroletti}, {Glanzman}, {Godfrey}, {Granot},
  {Greiner}, {Grenier}, {Grove}, {Guiriec}, {Gustafsson}, {Hadasch},
  {Hayashida}, {Hays}, {Horan}, {Hughes}, {J{\'o}hannesson}, {Johnson},
  {Johnson}, {Johnson}, {Kamae}, {Katagiri}, {Kataoka}, {Kn{\"o}dlseder},
  {Kocevski}, {Kuss}, {Lande}, {Latronico}, {Lee}, {Llena Garde}, {Longo},
  {Loparco}, {Lott}, {Lovellette}, {Lubrano}, {Makeev}, {Mazziotta},
  {McConville}, {McEnery}, {McGlynn}, {Mehault}, {M{\'e}sz{\'a}ros},
  {Michelson}, {Mizuno}, {Moiseev}, {Monte}, {Monzani}, {Moretti}, {Morselli},
  {Moskalenko}, {Murgia}, {Nakamori}, {Naumann-Godo}, {Nolan}, {Norris},
  {Nuss}, {Ohno}, {Ohsugi}, {Okumura}, {Omodei}, {Orlando}, {Ormes}, {Ozaki},
  {Paneque}, {Panetta}, {Parent}, {Pelassa}, {Pepe}, {Pesce-Rollins}, {Piron},
  {Porter}, {Primack}, {Rain{\`o}}, {Rando}, {Razzano}, {Razzaque}, {Reimer},
  {Reimer}, {Reyes}, {Ripken}, {Ritz}, {Romani}, {Roth}, {Sadrozinski},
  {Sanchez}, {Sander}, {Scargle}, {Schalk}, {Sgr{\`o}}, {Shaw}, {Siskind},
  {Smith}, {Spandre}, {Spinelli}, {Stecker}, {Strickman}, {Suson}, {Tajima},
  {Takahashi}, {Takahashi}, {Tanaka}, {Thayer}, {Thayer}, {Thompson},
  {Tibaldo}, {Torres}, {Tosti}, {Tramacere}, {Uchiyama}, {Usher},
  {Vandenbroucke}, {Vasileiou}, {Vilchez}, {Vitale}, {von Kienlin}, {Waite},
  {Wang}, {Wilson-Hodge}, {Winer}, {Wood}, {Yamazaki}, {Yang}, {Ylinen}, \&
  {Ziegler}}]{FermiEBL}
{Abdo}, A.~A., {Ackermann}, M., {Ajello}, M., {et~al.}  (Fermi/LAT collaboration) 2010, \apj, 723, 1082

\bibitem[{{Actis} {et~al.}(2011){Actis}, {Agnetta}, {Aharonian},
  {Akhperjanian}, {Aleksi{\'c}}, {Aliu}, {Allan}, {Allekotte}, {Antico},
  {Antonelli}, \& et~al.}]{2011ExA....32..193A}
{Actis}, M., {Agnetta}, G., {Aharonian}, F., {et~al.} 2011, Experimental
  Astronomy, 32, 193

\bibitem[{{Aharonian} {et~al.}(2003){Aharonian}, {Akhperjanian}, {Beilicke},
  {Bernl{\"o}hr}, {B{\"o}rst}, {Bojahr}, {Bolz}, {Coarasa}, {Contreras},
  {Cortina}, {Costamante}, {Denninghoff}, {Fonseca}, {Girma}, {G{\"o}tting},
  {Heinzelmann}, {Hermann}, {Heusler}, {Hofmann}, {Horns}, {Jung}, {Kankanyan},
  {Kestel}, {Kohnle}, {Konopelko}, {Kornmeyer}, {Kranich}, {Lampeitl}, {Lopez},
  {Lorenz}, {Lucarelli}, {Mang}, {Mazine}, {Meyer}, {Mirzoyan}, {Moralejo},
  {Ona-Wilhelmi}, {Panter}, {Plyasheshnikov}, {Prahl}, {P{\"u}hlhofer}, {de los
  Reyes}, {Rhode}, {Ripken}, {Rowell}, {Sahakian}, {Samorski}, {Schilling},
  {Siems}, {Sobzynska}, {Stamm}, {Tluczykont}, {Vitale}, {V{\"o}lk}, {Wiedner},
  \& {Wittek}}]{2003A&A...403..523A}
{Aharonian}, F., {Akhperjanian}, A., {Beilicke}, M., {et~al.} 2003, \aap, 403,
  523

\bibitem[{{Aharonian} {et~al.}(2009{\natexlab{a}}){Aharonian}, {Akhperjanian},
  {Anton}, {Barres de Almeida}, {Bazer-Bachi}, {Becherini}, {Behera}, {Benbow},
  {Bernl{\"o}hr}, {Boisson}, {Bochow}, {Borrel}, {Brion}, {Brucker}, {Brun},
  {B{\"u}hler}, {Bulik}, {B{\"u}sching}, {Boutelier}, {Chadwick},
  {Charbonnier}, {Chaves}, {Cheesebrough}, {Chounet}, {Clapson}, {Coignet},
  {Costamante}, {Dalton}, {Daniel}, {Davids}, {Degrange}, {Deil}, {Dickinson},
  {Djannati-Ata{\"i}}, {Domainko}, {O'C.~Drury}, {Dubois}, {Dubus}, {Dyks},
  {Dyrda}, {Egberts}, {Emmanoulopoulos}, {Espigat}, {Farnier}, {Feinstein},
  {Fiasson}, {F{\"o}rster}, {Fontaine}, {F{\"u}{\ss}ling}, {Gabici}, {Gallant},
  {G{\'e}rard}, {Giebels}, {Glicenstein}, {Gl{\"u}ck}, {Goret}, {G{\"o}hring},
  {Hauser}, {Hauser}, {Heinz}, {Heinzelmann}, {Henri}, {Hermann}, {Hinton},
  {Hoffmann}, {Hofmann}, {Holleran}, {Hoppe}, {Horns}, {Jacholkowska}, {de
  Jager}, {Jahn}, {Jung}, {Katarzy{\'n}ski}, {Katz}, {Kaufmann}, {Kendziorra},
  {Kerschhaggl}, {Khangulyan}, {Kh{\'e}lifi}, {Keogh}, {Klu{\'z}niak},
  {Kneiske}, {Komin}, {Kosack}, {Lamanna}, {Lenain}, {Lohse}, {Marandon},
  {Martin}, {Martineau-Huynh}, {Marcowith}, {Maurin}, {McComb}, {Medina},
  {Moderski}, {Monard}, {Moulin}, {Naumann-Godo}, {de Naurois}, {Nedbal},
  {Nekrassov}, {Niemiec}, {Nolan}, {Ohm}, {Olive}, {de O{\~n}a Wilhelmi},
  {Orford}, {Ostrowski}, {Panter}, {Paz Arribas}, {Pedaletti}, {Pelletier},
  {Petrucci}, {Pita}, {P{\"u}hlhofer}, {Punch}, {Quirrenbach}, {Raubenheimer},
  {Raue}, {Rayner}, {Renaud}, {Rieger}, {Ripken}, {Rob}, {Rosier-Lees},
  {Rowell}, {Rudak}, {Rulten}, {Ruppel}, {Sahakian}, {Santangelo},
  {Schlickeiser}, {Sch{\"o}ck}, {Schr{\"o}der}, {Schwanke}, {Schwarzburg},
  {Schwemmer}, {Shalchi}, {Sikora}, {Skilton}, {Sol}, {Spangler}, {Stawarz},
  {Steenkamp}, {Stegmann}, {Superina}, {Szostek}, {Tam}, {Tavernet}, {Terrier},
  {Tibolla}, {Tluczykont}, {van Eldik}, {Vasileiadis}, {Venter}, {Venter},
  {Vialle}, {Vincent}, {Vivier}, {V{\"o}lk}, {Volpe}, {Wagner}, {Ward},
  {Zdziarski}, \& {Zech}}]{2009A&A...502..749A}
{Aharonian}, F., {Akhperjanian}, A.~G., {Anton}, G., {et~al.} (H.E.S.S collaboration)
  2009{\natexlab{a}}, \aap, 502, 749

\bibitem[{{Aharonian} {et~al.}(2009{\natexlab{b}}){Aharonian}, {Akhperjanian},
  {Anton}, {Barres de Almeida}, {Bazer-Bachi}, {Becherini}, {Behera},
  {Bernl{\"o}hr}, {Boisson}, {Bochow}, \& et~al.}]{2155_2008}
{Aharonian}, F., {Akhperjanian}, A.~G., {Anton}, G., {et~al.} (H.E.S.S collaboration)
  2009{\natexlab{b}}, \apjl, 696, L150

\bibitem[{{Aharonian} {et~al.}(2005{\natexlab{a}}){Aharonian}, {Akhperjanian},
  {Aye}, {Bazer-Bachi}, {Beilicke}, {Benbow}, {Berge}, {Berghaus},
  {Bernl{\"o}hr}, {Boisson}, {Bolz}, {Braun}, {Breitling}, {Brown}, {Bussons
  Gordo}, {Chadwick}, {Chounet}, {Cornils}, {Costamante}, {Degrange},
  {Djannati-Ata{\"i}}, {O'C.~Drury}, {Dubus}, {Emmanoulopoulos}, {Espigat},
  {Feinstein}, {Fleury}, {Fontaine}, {Fuchs}, {Funk}, {Gallant}, {Giebels},
  {Gillessen}, {Glicenstein}, {Goret}, {Hadjichristidis}, {Hauser},
  {Heinzelmann}, {Henri}, {Hermann}, {Hinton}, {Hofmann}, {Holleran}, {Horns},
  {de Jager}, {Kh{\'e}lifi}, {Komin}, {Konopelko}, {Latham}, {Le Gallou},
  {Lemi{\`e}re}, {Lemoine}, {Leroy}, {Lohse}, {Marcowith}, {Masterson},
  {McComb}, {de Naurois}, {Nolan}, {Noutsos}, {Orford}, {Osborne}, {Ouchrif},
  {Panter}, {Pelletier}, {Pita}, {P{\"u}hlhofer}, {Punch}, {Raubenheimer},
  {Raue}, {Raux}, {Rayner}, {Redondo}, {Reimer}, {Reimer}, {Ripken}, {Rob},
  {Rolland}, {Rowell}, {Sahakian}, {Saug{\'e}}, {Schlenker}, {Schlickeiser},
  {Schuster}, {Schwanke}, {Siewert}, {Sol}, {Steenkamp}, {Stegmann},
  {Tavernet}, {Terrier}, {Th{\'e}oret}, {Tluczykont}, {Vasileiadis}, {Venter},
  {Vincent}, {V{\"o}lk}, \& {Wagner}}]{MrkHESS2004}
{Aharonian}, F., {Akhperjanian}, A.~G., {Aye}, K.-M., {et~al.} (H.E.S.S collaboration)
  2005{\natexlab{a}}, \aap, 437, 95

\bibitem[{{Aharonian} {et~al.}(2005{\natexlab{b}}){Aharonian}, {Akhperjanian},
  {Aye}, {Bazer-Bachi}, {Beilicke}, {Benbow}, {Berge}, {Berghaus},
  {Bernl{\"o}hr}, {Boisson}, {Bolz}, {Braun}, {Breitling}, {Brown}, {Bussons
  Gordo}, {Chadwick}, {Chounet}, {Cornils}, {Costamante}, {Degrange},
  {Djannati-Ata{\"i}}, {O'C.~Drury}, {Dubus}, {Emmanoulopoulos}, {Espigat},
  {Feinstein}, {Fleury}, {Fontaine}, {Fuchs}, {Funk}, {Gallant}, {Giebels},
  {Gillessen}, {Glicenstein}, {Goret}, {Hadjichristidis}, {Hauser},
  {Heinzelmann}, {Henri}, {Hermann}, {Hinton}, {Hofmann}, {Holleran}, {Horns},
  {de Jager}, {Kh{\'e}lifi}, {Komin}, {Konopelko}, {Latham}, {Le Gallou},
  {Lemi{\`e}re}, {Lemoine-Goumard}, {Leroy}, {Lohse}, {Martineau-Huynh},
  {Marcowith}, {Masterson}, {McComb}, {de Naurois}, {Nolan}, {Noutsos},
  {Orford}, {Osborne}, {Ouchrif}, {Panter}, {Pelletier}, {Pita},
  {P{\"u}hlhofer}, {Punch}, {Raubenheimer}, {Raue}, {Raux}, {Rayner},
  {Redondo}, {Reimer}, {Reimer}, {Ripken}, {Rob}, {Rolland}, {Rowell},
  {Sahakian}, {Saug{\'e}}, {Schlenker}, {Schlickeiser}, {Schuster}, {Schwanke},
  {Siewert}, {Sol}, {Steenkamp}, {Stegmann}, {Tavernet}, {Terrier},
  {Th{\'e}oret}, {Tluczykont}, {Vasileiadis}, {Venter}, {Vincent}, {V{\"o}lk},
  \& {Wagner}}]{2005begins}
{Aharonian}, F., {Akhperjanian}, A.~G., {Aye}, K.-M., {et~al.} (H.E.S.S collaboration)
  2005{\natexlab{b}}, \aap, 436, L17

\bibitem[{{Aharonian} {et~al.}(2005{\natexlab{c}}){Aharonian}, {Akhperjanian},
  {Aye}, {Bazer-Bachi}, {Beilicke}, {Benbow}, {Berge}, {Berghaus},
  {Bernl{\"o}hr}, {Bolz}, {Boisson}, {Borgmeier}, {Breitling}, {Brown},
  {Bussons Gordo}, {Chadwick}, {Chitnis}, {Chounet}, {Cornils}, {Costamante},
  {Degrange}, {Djannati-Ata{\"i}}, {Drury}, {Ergin}, {Espigat}, {Feinstein},
  {Fleury}, {Fontaine}, {Funk}, {Gallant}, {Giebels}, {Gillessen}, {Goret},
  {Guy}, {Hadjichristidis}, {Hauser}, {Heinzelmann}, {Henri}, {Hermann},
  {Hinton}, {Hofmann}, {Holleran}, {Horns}, {de Jager}, {Jung I.},
  {Kh{\'e}lifi}, {Komin}, {Konopelko}, {Latham}, {Le Gallou}, {Lemoine},
  {Lemi{\`e}re}, {Leroy}, {Lohse}, {Marcowith}, {Masterson}, {McComb}, {de
  Naurois}, {Nolan}, {Noutsos}, {Orford}, {Osborne}, {Ouchrif}, {Panter},
  {Pelletier}, {Pita}, {Pohl}, {P{\"u}hlhofer}, {Punch}, {Raubenheimer},
  {Raue}, {Raux}, {Rayner}, {Redondo}, {Reimer}, {Reimer}, {Ripken}, {Rivoal},
  {Rob}, {Rolland}, {Rowell}, {Sahakian}, {Saug{\'e}}, {Schlenker},
  {Schlickeiser}, {Schuster}, {Schwanke}, {Siewert}, {Sol}, {Steenkamp},
  {Stegmann}, {Tavernet}, {Th{\'e}oret}, {Tluczykont}, {van der Walt},
  {Vasileiadis}, {Vincent}, {Visser}, {V{\"o}lk}, \&
  {Wagner}}]{2005A&A...430..865A}
{Aharonian}, F., {Akhperjanian}, A.~G., {Aye}, K.-M., {et~al.} (H.E.S.S collaboration)
  2005{\natexlab{c}}, \aap, 430, 865

\bibitem[{{Aharonian} {et~al.}(2007{\natexlab{a}}){Aharonian}, {Akhperjanian},
  {Barres de Almeida}, {Bazer-Bachi}, {Behera}, {Beilicke}, {Benbow},
  {Bernl{\"o}hr}, {Boisson}, {Bolz}, {Borrel}, {Braun}, {Brion}, {Brown},
  {B{\"u}hler}, {Bulik}, {B{\"u}sching}, {Boutelier}, {Carrigan}, {Chadwick},
  {Chounet}, {Clapson}, {Coignet}, {Cornils}, {Costamante}, {Dalton},
  {Degrange}, {Dickinson}, {Djannati-Ata{\"i}}, {Domainko}, {O'C.~Drury},
  {Dubois}, {Dubus}, {Dyks}, {Egberts}, {Emmanoulopoulos}, {Espigat},
  {Farnier}, {Feinstein}, {Fiasson}, {F{\"o}rster}, {Fontaine}, {Funk},
  {F{\"u}{\ss}ling}, {Gallant}, {Giebels}, {Glicenstein}, {Gl{\"u}ck}, {Goret},
  {Hadjichristidis}, {Hauser}, {Hauser}, {Heinzelmann}, {Henri}, {Hermann},
  {Hinton}, {Hoffmann}, {Hofmann}, {Holleran}, {Hoppe}, {Horns},
  {Jacholkowska}, {de Jager}, {Jung}, {Katarzy{\'n}ski}, {Kendziorra},
  {Kerschhaggl}, {Kh{\'e}lifi}, {Keogh}, {Komin}, {Kosack}, {Lamanna},
  {Latham}, {Lemi{\`e}re}, {Lemoine-Goumard}, {Lenain}, {Lohse}, {Martin},
  {Martineau-Huynh}, {Marcowith}, {Masterson}, {Maurin}, {Maurin}, {McComb},
  {Moderski}, {Moulin}, {de Naurois}, {Nedbal}, {Nolan}, {Ohm}, {Olive}, {de
  O{\~n}a Wilhelmi}, {Orford}, {Osborne}, {Ostrowski}, {Panter}, {Pedaletti},
  {Pelletier}, {Petrucci}, {Pita}, {P{\"u}hlhofer}, {Punch}, {Ranchon},
  {Raubenheimer}, {Raue}, {Rayner}, {Renaud}, {Ripken}, {Rob}, {Rolland},
  {Rosier-Lees}, {Rowell}, {Rudak}, {Ruppel}, {Sahakian}, {Santangelo},
  {Schlickeiser}, {Sch{\"o}ck}, {Schr{\"o}der}, {Schwanke}, {Schwarzburg},
  {Schwemmer}, {Shalchi}, {Sol}, {Spangler}, {Stawarz}, {Steenkamp},
  {Stegmann}, {Superina}, {Tam}, {Tavernet}, {Terrier}, {van Eldik},
  {Vasileiadis}, {Venter}, {Vialle}, {Vincent}, {Vivier}, {V{\"o}lk}, {Volpe},
  {Wagner}, {Ward}, {Zdziarski}, \& {Zech}}]{0229}
{Aharonian}, F., {Akhperjanian}, A.~G., {Barres de Almeida}, U., {et~al.} (H.E.S.S collaboration)
  2007{\natexlab{a}}, \aap, 475, L9

\bibitem[{{Aharonian} {et~al.}(2007{\natexlab{b}}){Aharonian}, {Akhperjanian},
  {Barres de Almeida}, {Bazer-Bachi}, {Behera}, {Beilicke}, {Benbow},
  {Bernl{\"o}hr}, {Boisson}, {Bolz}, {Borrel}, {Braun}, {Brion}, {Brown},
  {B{\"u}hler}, {Bulik}, {B{\"u}sching}, {Boutelier}, {Carrigan}, {Chadwick},
  {Chounet}, {Clapson}, {Coignet}, {Cornils}, {Costamante}, {Dalton},
  {Degrange}, {Dickinson}, {Djannati-Ata{\"i}}, {Domainko}, {O'C.~Drury},
  {Dubois}, {Dubus}, {Dyks}, {Egberts}, {Emmanoulopoulos}, {Espigat},
  {Farnier}, {Feinstein}, {Fiasson}, {F{\"o}rster}, {Fontaine}, {Funk},
  {F{\"u}{\ss}ling}, {Gallant}, {Giebels}, {Glicenstein}, {Gl{\"u}ck}, {Goret},
  {Hadjichristidis}, {Hauser}, {Hauser}, {Heinzelmann}, {Henri}, {Hermann},
  {Hinton}, {Hoffmann}, {Hofmann}, {Holleran}, {Hoppe}, {Horns},
  {Jacholkowska}, {de Jager}, {Jung}, {Katarzy{\'n}ski}, {Kendziorra},
  {Kerschhaggl}, {Kh{\'e}lifi}, {Keogh}, {Komin}, {Kosack}, {Lamanna},
  {Latham}, {Lemi{\`e}re}, {Lemoine-Goumard}, {Lenain}, {Lohse}, {Martin},
  {Martineau-Huynh}, {Marcowith}, {Masterson}, {Maurin}, {Maurin}, {McComb},
  {Moderski}, {Moulin}, {de Naurois}, {Nedbal}, {Nolan}, {Ohm}, {Olive}, {de
  O{\~n}a Wilhelmi}, {Orford}, {Osborne}, {Ostrowski}, {Panter}, {Pedaletti},
  {Pelletier}, {Petrucci}, {Pita}, {P{\"u}hlhofer}, {Punch}, {Ranchon},
  {Raubenheimer}, {Raue}, {Rayner}, {Renaud}, {Ripken}, {Rob}, {Rolland},
  {Rosier-Lees}, {Rowell}, {Rudak}, {Ruppel}, {Sahakian}, {Santangelo},
  {Schlickeiser}, {Sch{\"o}ck}, {Schr{\"o}der}, {Schwanke}, {Schwarzburg},
  {Schwemmer}, {Shalchi}, {Sol}, {Spangler}, {Stawarz}, {Steenkamp},
  {Stegmann}, {Superina}, {Tam}, {Tavernet}, {Terrier}, {van Eldik},
  {Vasileiadis}, {Venter}, {Vialle}, {Vincent}, {Vivier}, {V{\"o}lk}, {Volpe},
  {Wagner}, {Ward}, {Zdziarski}, \& {Zech}}]{0347}
{Aharonian}, F., {Akhperjanian}, A.~G., {Barres de Almeida}, U., {et~al.} (H.E.S.S collaboration)
  2007{\natexlab{b}}, \aap, 473, L25

\bibitem[{{Aharonian} {et~al.}(2007{\natexlab{c}}){Aharonian}, {Akhperjanian},
  {Bazer-Bachi}, {Behera}, {Beilicke}, {Benbow}, {Berge}, {Bernl{\"o}hr},
  {Boisson}, {Bolz}, {Borrel}, {Boutelier}, {Braun}, {Brion}, {Brown},
  {B{\"u}hler}, {B{\"u}sching}, {Bulik}, {Carrigan}, {Chadwick}, {Clapson},
  {Chounet}, {Coignet}, {Cornils}, {Costamante}, {Degrange}, {Dickinson},
  {Djannati-Ata{\"i}}, {Domainko}, {Drury}, {Dubus}, {Dyks}, {Egberts},
  {Emmanoulopoulos}, {Espigat}, {Farnier}, {Feinstein}, {Fiasson},
  {F{\"o}rster}, {Fontaine}, {Funk}, {Funk}, {F{\"u}{\ss}ling}, {Gallant},
  {Giebels}, {Glicenstein}, {Gl{\"u}ck}, {Goret}, {Hadjichristidis}, {Hauser},
  {Hauser}, {Heinzelmann}, {Henri}, {Hermann}, {Hinton}, {Hoffmann}, {Hofmann},
  {Holleran}, {Hoppe}, {Horns}, {Jacholkowska}, {de Jager}, {Kendziorra},
  {Kerschhaggl}, {Kh{\'e}lifi}, {Komin}, {Kosack}, {Lamanna}, {Latham}, {Le
  Gallou}, {Lemi{\`e}re}, {Lemoine-Goumard}, {Lenain}, {Lohse}, {Martin},
  {Martineau-Huynh}, {Marcowith}, {Masterson}, {Maurin}, {McComb}, {Moderski},
  {Moulin}, {de Naurois}, {Nedbal}, {Nolan}, {Olive}, {Orford}, {Osborne},
  {Ostrowski}, {Panter}, {Pedaletti}, {Pelletier}, {Petrucci}, {Pita},
  {P{\"u}hlhofer}, {Punch}, {Ranchon}, {Raubenheimer}, {Raue}, {Rayner},
  {Renaud}, {Ripken}, {Rob}, {Rolland}, {Rosier-Lees}, {Rowell}, {Rudak},
  {Ruppel}, {Sahakian}, {Santangelo}, {Saug{\'e}}, {Schlenker}, {Schlickeiser},
  {Schr{\"o}der}, {Schwanke}, {Schwarzburg}, {Schwemmer}, {Shalchi}, {Sol},
  {Spangler}, {Stawarz}, {Steenkamp}, {Stegmann}, {Superina}, {Tam},
  {Tavernet}, {Terrier}, {van Eldik}, {Vasileiadis}, {Venter}, {Vialle},
  {Vincent}, {Vivier}, {V{\"o}lk}, {Volpe}, {Wagner}, {Ward}, \&
  {Zdziarski}}]{2007ApJ...664L..71A}
{Aharonian}, F., {Akhperjanian}, A.~G., {Bazer-Bachi}, A.~R., {et~al.} (H.E.S.S collaboration)
  2007{\natexlab{c}}, \apjl, 664, L71

\bibitem[{{Aharonian} {et~al.}(2006{\natexlab{a}}){Aharonian}, {Akhperjanian},
  {Bazer-Bachi}, {Beilicke}, {Benbow}, {Berge}, {Bernl{\"o}hr}, {Boisson},
  {Bolz}, {Borrel}, {Braun}, {Breitling}, {Brown}, {B{\"u}hler},
  {B{\"u}sching}, {Carrigan}, {Chadwick}, {Chounet}, {Cornils}, {Costamante},
  {Degrange}, {Dickinson}, {Djannati-Ata{\"i}}, {O'C.~Drury}, {Dubus},
  {Egberts}, {Emmanoulopoulos}, {Espigat}, {Feinstein}, {Ferrero}, {Fiasson},
  {Fontaine}, {Funk}, {Funk}, {Gallant}, {Giebels}, {Glicenstein}, {Goret},
  {Hadjichristidis}, {Hauser}, {Hauser}, {Heinzelmann}, {Henri}, {Hermann},
  {Hinton}, {Hofmann}, {Holleran}, {Horns}, {Jacholkowska}, {de Jager},
  {Kh{\'e}lifi}, {Komin}, {Konopelko}, {Kosack}, {Latham}, {Le Gallou},
  {Lemi{\`e}re}, {Lemoine-Goumard}, {Lohse}, {Martin}, {Martineau-Huynh},
  {Marcowith}, {Masterson}, {McComb}, {de Naurois}, {Nedbal}, {Nolan},
  {Noutsos}, {Orford}, {Osborne}, {Ouchrif}, {Panter}, {Pelletier}, {Pita},
  {P{\"u}hlhofer}, {Punch}, {Raubenheimer}, {Raue}, {Rayner}, {Reimer},
  {Reimer}, {Ripken}, {Rob}, {Rolland}, {Rowell}, {Sahakian}, {Saug{\'e}},
  {Schlenker}, {Schlickeiser}, {Schwanke}, {Sol}, {Spangler}, {Spanier},
  {Steenkamp}, {Stegmann}, {Superina}, {Tavernet}, {Terrier}, {Th{\'e}oret},
  {Tluczykont}, {van Eldik}, {Vasileiadis}, {Venter}, {Vincent}, {V{\"o}lk},
  {Wagner}, \& {Ward}}]{Crab}
{Aharonian}, F., {Akhperjanian}, A.~G., {Bazer-Bachi}, A.~R., {et~al.} (H.E.S.S collaboration)
  2006{\natexlab{a}}, \aap, 457, 899

\bibitem[{{Aharonian} {et~al.}(2006{\natexlab{b}}){Aharonian}, {Akhperjanian},
  {Bazer-Bachi}, {Beilicke}, {Benbow}, {Berge}, {Bernl{\"o}hr}, {Boisson},
  {Bolz}, {Borrel}, {Braun}, {Breitling}, {Brown}, {B{\"u}hler},
  {B{\"u}sching}, {Carrigan}, {Chadwick}, {Chounet}, {Cornils}, {Costamante},
  {Degrange}, {Dickinson}, {Djannati-Ata{\"i}}, {O'C.~Drury}, {Dubus},
  {Egberts}, {Emmanoulopoulos}, {Espigat}, {Feinstein}, {Ferrero}, {Fontaine},
  {Funk}, {Funk}, {Gallant}, {Giebels}, {Glicenstein}, {Goret},
  {Hadjichristidis}, {Hauser}, {Hauser}, {Heinzelmann}, {Henri}, {Hermann},
  {Hinton}, {Hofmann}, {Holleran}, {Horns}, {Jacholkowska}, {de Jager},
  {Kh{\'e}lifi}, {Komin}, {Konopelko}, {Latham}, {Le Gallou}, {Lemi{\`e}re},
  {Lemoine-Goumard}, {Lohse}, {Martin}, {Martineau-Huynh}, {Marcowith},
  {Masterson}, {McComb}, {de Naurois}, {Nedbal}, {Nolan}, {Noutsos}, {Orford},
  {Osborne}, {Ouchrif}, {Panter}, {Pelletier}, {Pita}, {P{\"u}hlhofer},
  {Punch}, {Raubenheimer}, {Raue}, {Rayner}, {Reimer}, {Reimer}, {Ripken},
  {Rob}, {Rolland}, {Rowell}, {Sahakian}, {Saug{\'e}}, {Schlenker},
  {Schlickeiser}, {Schwanke}, {Sol}, {Spangler}, {Spanier}, {Steenkamp},
  {Stegmann}, {Superina}, {Tavernet}, {Terrier}, {Th{\'e}oret}, {Tluczykont},
  {van Eldik}, {Vasileiadis}, {Venter}, {Vincent}, {V{\"o}lk}, {Wagner}, \&
  {Ward}}]{2356}
{Aharonian}, F., {Akhperjanian}, A.~G., {Bazer-Bachi}, A.~R., {et~al.} (H.E.S.S collaboration)
  2006{\natexlab{b}}, \aap, 455, 461

\bibitem[{{Aharonian} {et~al.}(2006{\natexlab{c}}){Aharonian}, {Akhperjanian},
  {Bazer-Bachi}, {Beilicke}, {Benbow}, {Berge}, {Bernl{\"o}hr}, {Boisson},
  {Bolz}, {Borrel}, {Braun}, {Breitling}, {Brown}, {Chadwick}, {Chounet},
  {Cornils}, {Costamante}, {Degrange}, {Dickinson}, {Djannati-Ata{\"i}},
  {Drury}, {Dubus}, {Emmanoulopoulos}, {Espigat}, {Feinstein}, {Fontaine},
  {Fuchs}, {Funk}, {Gallant}, {Giebels}, {Gillessen}, {Glicenstein}, {Goret},
  {Hadjichristidis}, {Hauser}, {Hauser}, {Heinzelmann}, {Henri}, {Hermann},
  {Hinton}, {Hofmann}, {Holleran}, {Horns}, {Jacholkowska}, {de Jager},
  {Kh{\'e}lifi}, {Klages}, {Komin}, {Konopelko}, {Latham}, {Le Gallou},
  {Lemi{\`e}re}, {Lemoine-Goumard}, {Leroy}, {Lohse}, {Martin},
  {Martineau-Huynh}, {Marcowith}, {Masterson}, {McComb}, {de Naurois}, {Nolan},
  {Noutsos}, {Orford}, {Osborne}, {Ouchrif}, {Panter}, {Pelletier}, {Pita},
  {P{\"u}hlhofer}, {Punch}, {Raubenheimer}, {Raue}, {Raux}, {Rayner}, {Reimer},
  {Reimer}, {Ripken}, {Rob}, {Rolland}, {Rowell}, {Sahakian}, {Saug{\'e}},
  {Schlenker}, {Schlickeiser}, {Schuster}, {Schwanke}, {Siewert}, {Sol},
  {Spangler}, {Steenkamp}, {Stegmann}, {Tavernet}, {Terrier}, {Th{\'e}oret},
  {Tluczykont}, {van Eldik}, {Vasileiadis}, {Venter}, {Vincent}, {V{\"o}lk}, \&
  {Wagner}}]{EBLAHA}
{Aharonian}, F., {Akhperjanian}, A.~G., {Bazer-Bachi}, A.~R., {et~al.} (H.E.S.S collaboration)
  2006{\natexlab{c}}, \nat, 440, 1018

\bibitem[{{Aharonian} {et~al.}(2005{\natexlab{d}}){Aharonian}, {Akhperjanian},
  {Bazer-Bachi}, {Beilicke}, {Benbow}, {Berge}, {Bernl{\"o}hr}, {Boisson},
  {Bolz}, {Borrel}, {Braun}, {Breitling}, {Brown}, {Chadwick}, {Chounet},
  {Cornils}, {Costamante}, {Degrange}, {Dickinson}, {Djannati-Ata{\"i}},
  {O'C.~Drury}, {Dubus}, {Emmanoulopoulos}, {Espigat}, {Feinstein}, {Fontaine},
  {Fuchs}, {Funk}, {Gallant}, {Giebels}, {Gillessen}, {Glicenstein}, {Goret},
  {Hadjichristidis}, {Hauser}, {Heinzelmann}, {Henri}, {Hermann}, {Hinton},
  {Hofmann}, {Holleran}, {Horns}, {Jacholkowska}, {de Jager}, {Kh{\'e}lifi},
  {Komin}, {Konopelko}, {Latham}, {Le Gallou}, {Lemi{\`e}re},
  {Lemoine-Goumard}, {Leroy}, {Lohse}, {Martin}, {Martineau-Huynh},
  {Marcowith}, {Masterson}, {McComb}, {de Naurois}, {Nolan}, {Noutsos},
  {Orford}, {Osborne}, {Ouchrif}, {Panter}, {Pelletier}, {Pita},
  {P{\"u}hlhofer}, {Punch}, {Raubenheimer}, {Raue}, {Raux}, {Rayner}, {Reimer},
  {Reimer}, {Ripken}, {Rob}, {Rolland}, {Rowell}, {Sahakian}, {Saug{\'e}},
  {Schlenker}, {Schlickeiser}, {Schuster}, {Schwanke}, {Siewert}, {Sol},
  {Spangler}, {Steenkamp}, {Stegmann}, {Tavernet}, {Terrier}, {Th{\'e}oret},
  {Tluczykont}, {Vasileiadis}, {Venter}, {Vincent}, {V{\"o}lk}, \&
  {Wagner}}]{2005A&A...442..895A}
{Aharonian}, F., {Akhperjanian}, A.~G., {Bazer-Bachi}, A.~R., {et~al.} (H.E.S.S collaboration)
  2005{\natexlab{d}}, \aap, 442, 895

\bibitem[{{Aharonian} {et~al.}(2007{\natexlab{d}}){Aharonian}, {Akhperjanian},
  {Bazer-Bachi}, {Beilicke}, {Benbow}, {Berge}, {Bernl{\"o}hr}, {Boisson},
  {Bolz}, {Borrel}, {Braun}, {Brion}, {Brown}, {B{\"u}hler}, {B{\"u}sching},
  {Boutelier}, {Carrigan}, {Chadwick}, {Chounet}, {Coignet}, {Cornils},
  {Costamante}, {Degrange}, {Dickinson}, {Djannati-Ata{\"i}}, {O'C.~Drury},
  {Dubus}, {Egberts}, {Emmanoulopoulos}, {Espigat}, {Farnier}, {Feinstein},
  {Ferrero}, {Fiasson}, {Fontaine}, {Funk}, {Funk}, {F{\"u}{\ss}ling},
  {Gallant}, {Giebels}, {Glicenstein}, {Gl{\"u}ck}, {Goret}, {Hadjichristidis},
  {Hauser}, {Hauser}, {Heinzelmann}, {Henri}, {Hermann}, {Hinton}, {Hoffmann},
  {Hofmann}, {Holleran}, {Hoppe}, {Horns}, {Jacholkowska}, {de Jager},
  {Kendziorra}, {Kerschhaggl}, {Kh{\'e}lifi}, {Komin}, {Kosack}, {Lamanna},
  {Latham}, {Le Gallou}, {Lemi{\`e}re}, {Lemoine-Goumard}, {Lohse}, {Martin},
  {Martineau-Huynh}, {Marcowith}, {Masterson}, {Maurin}, {McComb}, {Moulin},
  {de Naurois}, {Nedbal}, {Nolan}, {Noutsos}, {Olive}, {Orford}, {Osborne},
  {Panter}, {Pelletier}, {Petrucci}, {Pita}, {P{\"u}hlhofer}, {Punch},
  {Ranchon}, {Raubenheimer}, {Raue}, {Rayner}, {Ripken}, {Rob}, {Rolland},
  {Rosier-Lees}, {Rowell}, {Sahakian}, {Santangelo}, {Saug{\'e}}, {Schlenker},
  {Schlickeiser}, {Schr{\"o}der}, {Schwanke}, {Schwarzburg}, {Schwemmer},
  {Shalchi}, {Sol}, {Spangler}, {Spanier}, {Steenkamp}, {Stegmann}, {Superina},
  {Tam}, {Tavernet}, {Terrier}, {Tluczykont}, {van Eldik}, {Vasileiadis},
  {Venter}, {Vialle}, {Vincent}, {V{\"o}lk}, {Wagner}, \& {Ward}}]{1101}
{Aharonian}, F., {Akhperjanian}, A.~G., {Bazer-Bachi}, A.~R., {et~al.} (H.E.S.S collaboration)
  2007{\natexlab{d}}, \aap, 470, 475

\bibitem[{{Aharonian}(2000)}]{2000NewA....5..377A}
{Aharonian}, F.~A. 2000, \na, 5, 377

\bibitem[{{Aharonian} {et~al.}(1999){Aharonian}, {Akhperjanian}, {Barrio},
  {Bernl{\"o}hr}, {Bojahr}, {Calle}, {Contreras}, {Cortina}, {Daum}, {Deckers},
  {Denninghoff}, {Fonseca}, {Gonzalez}, {Heinzelmann}, {Hemberger}, {Hermann},
  {He{\ss}}, {Heusler}, {Hofmann}, {Hohl}, {Horns}, {Ibarra}, {Kankanyan},
  {Kettler}, {K{\"o}hler}, {Konopelko}, {Kornmeyer}, {Kestel}, {Kranich},
  {Krawczynski}, {Lampeitl}, {Lindner}, {Lorenz}, {Magnussen}, {Meyer},
  {Mirzoyan}, {Moralejo}, {Padilla}, {Panter}, {Petry}, {Plaga},
  {Plyasheshnikov}, {Prahl}, {P{\"u}hlhofer}, {Rauterberg}, {Renault}, {Rhode},
  {R{\"o}hring}, {Sahakian}, {Samorski}, {Schmele}, {Schr{\"o}der}, {Stamm},
  {V{\"o}lk}, {Wiebel-Sooth}, {Wiedner}, {Willmer}, \&
  {Wittek}}]{1999A&A...349...11A}
{Aharonian}, F.~A., {Akhperjanian}, A.~G., {Barrio}, J.~A., {et~al.} 1999,
  \aap, 349, 11

\bibitem[{{Band} \& {Grindlay}(1985)}]{1985ApJ...298..128B}
{Band}, D.~L. \& {Grindlay}, J.~E. 1985, \apj, 298, 128

\bibitem[{{Beall} \& {Bednarek}(1999)}]{1999ApJ...510..188B}
{Beall}, J.~H. \& {Bednarek}, W. 1999, \apj, 510, 188

\bibitem[{{Bernlohr}(2000)}]{Bernlohr00}
{Bernlohr}, K. 2000, Astroparticle Physics, 12, 255

\bibitem[{{Bernl{\"o}hr} {et~al.}(2003){Bernl{\"o}hr}, {Carrol}, {Cornils},
  {Elfahem}, {Espigat}, {Gillessen}, {Heinzelmann}, {Hermann}, {Hofmann},
  {Horns}, {Jung}, {Kankanyan}, {Katona}, {Khelifi}, {Krawczynski}, {Panter},
  {Punch}, {Rayner}, {Rowell}, {Tluczykont}, \& {van Staa}}]{Bernlohr}
{Bernl{\"o}hr}, K., {Carrol}, O., {Cornils}, R., {et~al.} 2003, Astroparticle
  Physics, 20, 111

\bibitem[{{de Naurois} \& {Rolland}(2009)}]{MathieuANA}
{de Naurois}, M. \& {Rolland}, L. 2009, Astroparticle Physics, 32, 231

\bibitem[{{Dole} {et~al.}(2006){Dole}, {Lagache}, {Puget}, {Caputi},
  {Fern{\'a}ndez-Conde}, {Le Floc'h}, {Papovich}, {P{\'e}rez-Gonz{\'a}lez},
  {Rieke}, \& {Blaylock}}]{SummaryObs2}
{Dole}, H., {Lagache}, G., {Puget}, J.-L., {et~al.} 2006, \aap, 451, 417

\bibitem[{{Dom{\'{\i}}nguez} {et~al.}(2011){Dom{\'{\i}}nguez}, {Primack},
  {Rosario}, {Prada}, {Gilmore}, {Faber}, {Koo}, {Somerville},
  {P{\'e}rez-Torres}, {P{\'e}rez-Gonz{\'a}lez}, {Huang}, {Davis},
  {Guhathakurta}, {Barmby}, {Conselice}, {Lozano}, {Newman}, \&
  {Cooper}}]{Dominguez}
{Dom{\'{\i}}nguez}, A., {Primack}, J.~R., {Rosario}, D.~J., {et~al.} 2011,
  \mnras, 410, 2556

\bibitem[{{Falomo} {et~al.}(1987){Falomo}, {Maraschi}, {Treves}, \&
  {Tanzi}}]{1987ApJ...318L..39F}
{Falomo}, R., {Maraschi}, L., {Treves}, A., \& {Tanzi}, E.~G. 1987, \apjl, 318,
  L39

\bibitem[{{Falomo} {et~al.}(1993){Falomo}, {Pesce}, \&
  {Treves}}]{1993ApJ...411L..63F}
{Falomo}, R., {Pesce}, J.~E., \& {Treves}, A. 1993, \apjl, 411, L63

\bibitem[{{Fazio} {et~al.}(2004){Fazio}, {Ashby}, {Barmby}, {Hora}, {Huang},
  {Pahre}, {Wang}, {Willner}, {Arendt}, {Moseley}, {Brodwin}, {Eisenhardt},
  {Stern}, {Tollestrup}, \& {Wright}}]{2004ApJS..154...39F}
{Fazio}, G.~G., {Ashby}, M.~L.~N., {Barmby}, P., {et~al.} 2004, \apjs, 154, 39

\bibitem[{{Franceschini} {et~al.}(2008){Franceschini}, {Rodighiero}, \&
  {Vaccari}}]{Fr08}
{Franceschini}, A., {Rodighiero}, G., \& {Vaccari}, M. 2008, \aap, 487, 837

\bibitem[{{Funk} {et~al.}(2004){Funk}, {Hermann}, {Hinton}, {Berge},
  {Bernl{\"o}hr}, {Hofmann}, {Nayman}, {Toussenel}, \& {Vincent}}]{Funk}
{Funk}, S., {Hermann}, G., {Hinton}, J., {et~al.} 2004, Astroparticle Physics,
  22, 285

\bibitem[{{Georganopoulos} {et~al.}(2010){Georganopoulos}, {Finke}, \&
  {Reyes}}]{Georganopoulos}
{Georganopoulos}, M., {Finke}, J.~D., \& {Reyes}, L.~C. 2010, \apjl, 714, L157

\bibitem[{{Gilmore} {et~al.}(2012){Gilmore}, {Somerville}, {Primack}, \&
  {Dom{\'{\i}}nguez}}]{Gilmore}
{Gilmore}, R.~C., {Somerville}, R.~S., {Primack}, J.~R., \& {Dom{\'{\i}}nguez},
  A. 2012, \mnras, 422, 3189

\bibitem[{{Gould} \& {Schr{\'e}der}(1967)}]{1967PhRv..155.1408G}
{Gould}, R.~J. \& {Schr{\'e}der}, G.~P. 1967, Physical Review, 155, 1408

\bibitem[{{Hauser} \& {Dwek}(2001)}]{SummaryObs1}
{Hauser}, M.~G. \& {Dwek}, E. 2001, \araa, 39, 249

\bibitem[{{Herterich}(1974)}]{1974Natur.250..311H}
{Herterich}, K. 1974, \nat, 250, 311

\bibitem[{{H.E.S.S.~collaboration, Abramowski}
  {et~al.}(2012){H.E.S.S.~collaboration, Abramowski}, {Acero}, {Aharonian},
  {Akhperjanian}, {Anton}, {Balzer}, {Barnacka}, {Barres de Almeida},
  {Becherini}, \& et~al.}]{2012arXiv1201.4135H}
{H.E.S.S.~collaboration, Abramowski}, A., {Acero}, F., {Aharonian}, F.,
  {et~al.} 2012, \aap, 539, A149

\bibitem[{{H.E.S.S.~collaboration, Abramowski}
  {et~al.}(2011){H.E.S.S.~collaboration, Abramowski}, {Acero}, {Aharonian},
  {Akhperjanian}, {Anton}, {Barnacka}, {Barres de Almeida}, {Bazer-Bachi},
  {Becherini}, \& et~al.}]{2011A&A...533A.110H}
{H.E.S.S.~collaboration, Abramowski}, A., {Acero}, F., {Aharonian}, F.,
  {et~al.} 2011, \aap, 533, A110

\bibitem[{{H.E.S.S.~collaboration, Abramowski}
  {et~al.}(2010{\natexlab{a}}){H.E.S.S.~collaboration, Abramowski}, {Acero},
  {Aharonian}, {Akhperjanian}, {Anton}, {Barres de Almeida}, {Bazer-Bachi},
  {Becherini}, {Behera}, {Benbow}, {Bernl{\"o}hr}, {Bochow}, {Boisson},
  {Bolmont}, {Borrel}, {Brucker}, {Brun}, {Brun}, {B{\"u}hler}, {Bulik},
  {B{\"u}sching}, {Boutelier}, {Chadwick}, {Charbonnier}, {Chaves},
  {Cheesebrough}, {Conrad}, {Chounet}, {Clapson}, {Coignet}, {Costamante},
  {Dalton}, {Daniel}, {Davids}, {Degrange}, {Deil}, {Dickinson},
  {Djannati-Ata{\"i}}, {Domainko}, {O'C.~Drury}, {Dubois}, {Dubus}, {Dyks},
  {Dyrda}, {Egberts}, {Eger}, {Espigat}, {Fallon}, {Farnier}, {Fegan},
  {Feinstein}, {Fernandes}, {Fiasson}, {F{\"o}rster}, {Fontaine},
  {F{\"u}{\ss}ling}, {Gabici}, {Gallant}, {G{\'e}rard}, {Gerbig}, {Giebels},
  {Glicenstein}, {Gl{\"u}ck}, {Goret}, {G{\"o}ring}, {Hampf}, {Hauser},
  {Heinz}, {Heinzelmann}, {Henri}, {Hermann}, {Hinton}, {Hoffmann}, {Hofmann},
  {Hofverberg}, {Holleran}, {Hoppe}, {Horns}, {Jacholkowska}, {de Jager},
  {Jahn}, {Jung}, {Katarzy{\'n}ski}, {Katz}, {Kaufmann}, {Kerschhaggl},
  {Khangulyan}, {Kh{\'e}lifi}, {Keogh}, {Klochkov}, {Klu{\v z}niak}, {Kneiske},
  {Komin}, {Kosack}, {Kossakowski}, {Lamanna}, {Lenain}, {Lohse}, {Lu},
  {Marandon}, {Marcowith}, {Masbou}, {Maurin}, {McComb}, {Medina},
  {M{\'e}hault}, {Moderski}, {Moulin}, {Naumann-Godo}, {de Naurois}, {Nedbal},
  {Nekrassov}, {Nguyen}, {Nicholas}, {Niemiec}, {Nolan}, {Ohm}, {Olive}, {de
  O{\~n}a Wilhelmi}, {Opitz}, {Orford}, {Ostrowski}, {Panter}, {Paz Arribas},
  {Pedaletti}, {Pelletier}, {Petrucci}, {Pita}, {P{\"u}hlhofer}, {Punch},
  {Quirrenbach}, {Raubenheimer}, {Raue}, {Rayner}, {Reimer}, {Renaud}, {de Los
  Reyes}, {Rieger}, {Ripken}, {Rob}, {Rosier-Lees}, {Rowell}, {Rudak},
  {Rulten}, {Ruppel}, {Ryde}, {Sahakian}, {Santangelo}, {Schlickeiser},
  {Sch{\"o}ck}, {Sch{\"o}nwald}, {Schwanke}, {Schwarzburg}, {Schwemmer},
  {Shalchi}, {Sushch}, {Sikora}, {Skilton}, {Sol}, {Stawarz}, {Steenkamp},
  {Stegmann}, {Stinzing}, {Szostek}, {Tam}, {Tavernet}, {Terrier}, {Tibolla},
  {Tluczykont}, {Valerius}, {van Eldik}, {Vasileiadis}, {Venter}, {Venter},
  {Vialle}, {Viana}, {Vincent}, {Vivier}, {V{\"o}lk}, {Volpe}, {Vorobiov},
  {Wagner}, {Ward}, {Zdziarski}, {Zech}, \& {Zechlin}}]{2010A&A...516A..56H}
{H.E.S.S.~collaboration, Abramowski}, A., {Acero}, F., {Aharonian}, F.,
  {et~al.} 2010{\natexlab{a}}, \aap, 516, A56

\bibitem[{{H.E.S.S.~collaboration, Abramowski}
  {et~al.}(2010{\natexlab{b}}){H.E.S.S.~collaboration, Abramowski}, {Acero},
  {Aharonian}, {Akhperjanian}, {Anton}, {Barres de Almeida}, {Bazer-Bachi},
  {Becherini}, {Benbow}, {Bernl{\"o}hr}, {Bochow}, {Boisson}, {Bolmont},
  {Borrel}, {Brucker}, {Brun}, {Brun}, {B{\"u}hler}, {Bulik}, {B{\"u}sching},
  {Boutelier}, {Chadwick}, {Charbonnier}, {Chaves}, {Cheesebrough}, {Chounet},
  {Clapson}, {Coignet}, {Conrad}, {Costamante}, {Dalton}, {Daniel}, {Davids},
  {Degrange}, {Deil}, {Dickinson}, {Djannati-Ata{\"i}}, {Domainko},
  {O'C.~Drury}, {Dubois}, {Dubus}, {Dyks}, {Dyrda}, {Egberts}, {Eger},
  {Espigat}, {Fallon}, {Farnier}, {Fegan}, {Feinstein}, {Fernandes}, {Fiasson},
  {F{\"o}rster}, {Fontaine}, {F{\"u}{\ss}ling}, {Gabici}, {Gallant},
  {G{\'e}rard}, {Gerbig}, {Giebels}, {Glicenstein}, {Gl{\"u}ck}, {Goret},
  {G{\"o}ring}, {Hampf}, {Hauser}, {Heinz}, {Heinzelmann}, {Henri}, {Hermann},
  {Hinton}, {Hoffmann}, {Hofmann}, {Hofverberg}, {Holleran}, {Hoppe}, {Horns},
  {Jacholkowska}, {de Jager}, {Jahn}, {Jung}, {Katarzy{\'n}ski}, {Katz},
  {Kaufmann}, {Kerschhaggl}, {Khangulyan}, {Kh{\'e}lifi}, {Keogh}, {Klochkov},
  {Klu{\'z}niak}, {Kneiske}, {Komin}, {Kosack}, {Kossakowski}, {Lamanna},
  {Lenain}, {Lohse}, {Lu}, {Marandon}, {Marcowith}, {Masbou}, {Maurin},
  {McComb}, {Medina}, {M{\'e}hault}, {Moderski}, {Moulin}, {Naumann-Godo}, {de
  Naurois}, {Nedbal}, {Nekrassov}, {Nguyen}, {Nicholas}, {Niemiec}, {Nolan},
  {Ohm}, {Olive}, {de O{\~n}a Wilhelmi}, {Opitz}, {Orford}, {Ostrowski},
  {Panter}, {Paz Arribas}, {Pedaletti}, {Pelletier}, {Petrucci}, {Pita},
  {P{\"u}hlhofer}, {Punch}, {Quirrenbach}, {Raubenheimer}, {Raue}, {Rayner},
  {Reimer}, {Renaud}, {de los Reyes}, {Rieger}, {Ripken}, {Rob}, {Rosier-Lees},
  {Rowell}, {Rudak}, {Rulten}, {Ruppel}, {Ryde}, {Sahakian}, {Santangelo},
  {Schlickeiser}, {Sch{\"o}ck}, {Sch{\"o}nwald}, {Schwanke}, {Schwarzburg},
  {Schwemmer}, {Shalchi}, {Sushch}, {Sikora}, {Skilton}, {Sol}, {Stawarz},
  {Steenkamp}, {Stegmann}, {Stinzing}, {Superina}, {Szostek}, {Tam},
  {Tavernet}, {Terrier}, {Tibolla}, {Tluczykont}, {Valerius}, {van Eldik},
  {Vasileiadis}, {Venter}, {Venter}, {Vialle}, {Viana}, {Vincent}, {Vivier},
  {V{\"o}lk}, {Volpe}, {Vorobiov}, {Wagner}, {Ward}, {Zdziarski}, {Zech}, \&
  {Zechlin}}]{2155_2010}
{H.E.S.S.~collaboration, Abramowski}, A., {Acero}, F., {Aharonian}, F.,
  {et~al.} 2010{\natexlab{b}}, \aap, 520, A83

\bibitem[{{H.E.S.S.~collaboration, Acero}
  {et~al.}(2010){H.E.S.S.~collaboration, Acero}, {Aharonian}, {Akhperjanian},
  {Anton}, {Barres de Almeida}, {Bazer-Bachi}, {Becherini}, {Behera}, {Benbow},
  {Bernl{\"o}hr}, {Bochow}, {Boisson}, {Bolmont}, {Borrel}, {Brucker}, {Brun},
  {Brun}, {B{\"u}hler}, {Bulik}, {B{\"u}sching}, {Boutelier}, {Chadwick},
  {Charbonnier}, {Chaves}, {Cheesebrough}, {Chounet}, {Clapson}, {Coignet},
  {Costamante}, {Dalton}, {Daniel}, {Davids}, {Degrange}, {Deil}, {Dickinson},
  {Djannati-Ata{\"i}}, {Domainko}, {O'C.~Drury}, {Dubois}, {Dubus}, {Dyks},
  {Dyrda}, {Egberts}, {Eger}, {Espigat}, {Fallon}, {Farnier}, {Fegan},
  {Feinstein}, {Fiasson}, {F{\"o}rster}, {Fontaine}, {F{\"u}{\ss}ling},
  {Gabici}, {Gallant}, {G{\'e}rard}, {Gerbig}, {Giebels}, {Glicenstein},
  {Gl{\"u}ck}, {Goret}, {G{\"o}ring}, {Hauser}, {Heinz}, {Heinzelmann},
  {Henri}, {Hermann}, {Hinton}, {Hoffmann}, {Hofmann}, {Hofverberg},
  {Holleran}, {Hoppe}, {Horns}, {Jacholkowska}, {de Jager}, {Jahn}, {Jung},
  {Katarzy{\'n}ski}, {Katz}, {Kaufmann}, {Kerschhaggl}, {Khangulyan},
  {Kh{\'e}lifi}, {Keogh}, {Klochkov}, {Klu{\'z}niak}, {Kneiske}, {Komin},
  {Kosack}, {Kossakowski}, {Lamanna}, {Lenain}, {Lohse}, {Marandon},
  {Martineau-Huynh}, {Marcowith}, {Masbou}, {Maurin}, {McComb}, {Medina},
  {M{\'e}hault}, {Moderski}, {Moulin}, {Naumann-Godo}, {de Naurois}, {Nedbal},
  {Nekrassov}, {Nicholas}, {Niemiec}, {Nolan}, {Ohm}, {Olive}, {de O{\~n}a
  Wilhelmi}, {Orford}, {Ostrowski}, {Panter}, {Paz Arribas}, {Pedaletti},
  {Pelletier}, {Petrucci}, {Pita}, {P{\"u}hlhofer}, {Punch}, {Quirrenbach},
  {Raubenheimer}, {Raue}, {Rayner}, {Renaud}, {Rieger}, {Ripken}, {Rob},
  {Rosier-Lees}, {Rowell}, {Rudak}, {Rulten}, {Ruppel}, {Sahakian},
  {Santangelo}, {Schlickeiser}, {Sch{\"o}ck}, {Schwanke}, {Schwarzburg},
  {Schwemmer}, {Shalchi}, {Sikora}, {Skilton}, {Sol}, {Stawarz}, {Steenkamp},
  {Stegmann}, {Stinzing}, {Superina}, {Szostek}, {Tam}, {Tavernet}, {Terrier},
  {Tibolla}, {Tluczykont}, {van Eldik}, {Vasileiadis}, {Venter}, {Venter},
  {Vialle}, {Vincent}, {Vivier}, {V{\"o}lk}, {Volpe}, {Wagner}, {Ward},
  {Zdziarski}, \& {Zech}}]{2005}
{H.E.S.S.~collaboration, Acero}, F., {Aharonian}, F., {Akhperjanian}, A.~G.,
  {et~al.} 2010, \aap, 511, A52

\bibitem[{{Hinton}(2004)}]{Hinton2004}
{Hinton}, J.~A. 2004, \nar, 48, 331

\bibitem[{{Jauch} \& {Rohrlich}(1976)}]{1976tper.book.....J}
{Jauch}, J.~M. \& {Rohrlich}, F. 1976, {The theory of photons and electrons.
The relativistic quantum field theory of charged particles with spin
one-half}

\bibitem[{{Jelley}(1966)}]{1966PhRvL..16..479J}
{Jelley}, J.~V. 1966, Physical Review Letters, 16, 479

\bibitem[{{Jones} {et~al.}(2009){Jones}, {Read}, {Saunders}, {Colless},
  {Jarrett}, {Parker}, {Fairall}, {Mauch}, {Sadler}, {Watson}, {Burton},
  {Campbell}, {Cass}, {Croom}, {Dawe}, {Fiegert}, {Frankcombe}, {Hartley},
  {Huchra}, {James}, {Kirby}, {Lahav}, {Lucey}, {Mamon}, {Moore}, {Peterson},
  {Prior}, {Proust}, {Russell}, {Safouris}, {Wakamatsu}, {Westra}, \&
  {Williams}}]{2009MNRAS.399..683J}
{Jones}, D.~H., {Read}, M.~A., {Saunders}, W., {et~al.} 2009, \mnras, 399, 683

\bibitem[{{Katarzy{\'n}ski} {et~al.}(2006){Katarzy{\'n}ski}, {Ghisellini},
  {Tavecchio}, {Gracia}, \& {Maraschi}}]{2006MNRAS.368L..52K}
{Katarzy{\'n}ski}, K., {Ghisellini}, G., {Tavecchio}, F., {Gracia}, J., \&
  {Maraschi}, L. 2006, \mnras, 368, L52

\bibitem[{{Lefa} {et~al.}(2011){Lefa}, {Rieger}, \&
  {Aharonian}}]{2011ApJ...740...64L}
{Lefa}, E., {Rieger}, F.~M., \& {Aharonian}, F. 2011, \apj, 740, 64

\bibitem[{Li \& Ma(1983)}]{LiMa}
Li, T.-P. \& Ma, Y.-Q. 1983, \apj, 272, 317

\bibitem[{{Madau} \& {Pozzetti}(2000)}]{2000MNRAS.312L...9M}
{Madau}, P. \& {Pozzetti}, L. 2000, \mnras, 312, L9

\bibitem[{{Mannheim}(1993)}]{1993A&A...269...67M}
{Mannheim}, K. 1993, \aap, 269, 67

\bibitem[{{Mazin} \& {Raue}(2007)}]{EBLMaRa}
{Mazin}, D. \& {Raue}, M. 2007, \aap, 471, 439

\bibitem[{{Meyer} {et~al.}(2010){Meyer}, {Horns}, \&
  {Zechlin}}]{2010A&A...523A...2M}
{Meyer}, M., {Horns}, D., \& {Zechlin}, H.-S. 2010, \aap, 523, A2

\bibitem[{{Meyer} {et~al.}(2012){Meyer}, {Raue}, {Mazin}, \& {Horns}}]{Meyer12}
{Meyer}, M., {Raue}, M., {Mazin}, D., \& {Horns}, D. 2012, arXiv:1202.2867

\bibitem[{{Nikishov}(1962)}]{REF::NIKISHOV::JETP1962}
{Nikishov}, A.~I. 1962, Soviet Physics JETP, 14, 393

\bibitem[{{Ohm} {et~al.}(2009){Ohm}, {van Eldik}, \& {Egberts}}]{TMVAHD}
{Ohm}, S., {van Eldik}, C., \& {Egberts}, K. 2009, Astroparticle Physics, 31,
  383

\bibitem[{{Orr} {et~al.}(2011){Orr}, {Krennrich}, \& {Dwek}}]{Orr}
{Orr}, M.~R., {Krennrich}, F., \& {Dwek}, E. 2011, \apj, 733, 77

\bibitem[{{Piron} {et~al.}(2001){Piron}, {Djannati-Atai}, {Punch}, {Tavernet},
  {Barrau}, {Bazer-Bachi}, {Chounet}, {Debiais}, {Degrange}, {Dezalay},
  {Espigat}, {Fabre}, {Fleury}, {Fontaine}, {Goret}, {Gouiffes}, {Khelifi},
  {Malet}, {Masterson}, {Mohanty}, {Nuss}, {Renault}, {Rivoal}, {Rob}, \&
  {Vorobiov}}]{2001A&A...374..895P}
{Piron}, F., {Djannati-Atai}, A., {Punch}, M., {et~al.} 2001, \aap, 374, 895

\bibitem[{{Punch} {et~al.}(1992){Punch}, {Akerlof}, {Cawley}, {Chantell},
  {Fegan}, {Fennell}, {Gaidos}, {Hagan}, {Hillas}, {Jiang}, {Kerrick}, {Lamb},
  {Lawrence}, {Lewis}, {Meyer}, {Mohanty}, {O'Flaherty}, {Reynolds}, {Rovero},
  {Schubnell}, {Sembroski}, {Weekes}, \& {Wilson}}]{Punch92}
{Punch}, M., {Akerlof}, C.~W., {Cawley}, M.~F., {et~al.} 1992, \nat, 358, 477

\bibitem[{{Raue} \& {Mazin}(2010)}]{2010APh....34..245R}
{Raue}, M. \& {Mazin}, D. 2010, Astroparticle Physics, 34, 245

\bibitem[{{Reimer}(2007)}]{Reimer2007}
{Reimer}, A. 2007, \apj, 665, 1023

\bibitem[{{Remillard} {et~al.}(1989){Remillard}, {Tuohy}, {Brissenden},
  {Buckley}, {Schwartz}, {Feigelson}, \& {Tapia}}]{1989ApJ...345..140R}
{Remillard}, R.~A., {Tuohy}, I.~R., {Brissenden}, R.~J.~V., {et~al.} 1989,
  \apj, 345, 140

\bibitem[{{Schachter} {et~al.}(1993){Schachter}, {Stocke}, {Perlman}, {Elvis},
  {Remillard}, {Granados}, {Luu}, {Huchra}, {Humphreys}, {Urry}, \&
  {Wallin}}]{1993ApJ...412..541S}
{Schachter}, J.~F., {Stocke}, J.~T., {Perlman}, E., {et~al.} 1993, \apj, 412,
  541

\bibitem[{{Sinervo}(2003)}]{Sinervo}
{Sinervo}, P. 2003, in Statistical Problems in Particle Physics, Astrophysics,
  and Cosmology, ed. {L.~Lyons, R.~Mount, \& R.~Reitmeyer}, 122

\bibitem[{{Stecker} {et~al.}(1992){Stecker}, {de Jager}, \&
{Salamon}}]{1992ApJ...390L..49S}
{Stecker}, F.~W., {de Jager}, O.~C., \& {Salamon}, M.~H. 1992, \apjl, 390, L49

\bibitem[{{The Fermi-LAT Collaboration}(2012)}]{2012arXiv1211.1671T}
{The Fermi-LAT Collaboration}. 2012, arXiv:1211.1671

\bibitem[{{Ulrich} {et~al.}(1975){Ulrich}, {Kinman}, {Lynds}, {Rieke}, \&
  {Ekers}}]{1975ApJ...198..261U}
{Ulrich}, M.-H., {Kinman}, T.~D., {Lynds}, C.~R., {Rieke}, G.~H., \& {Ekers},
  R.~D. 1975, \apj, 198, 261

\bibitem[{{Vaughan} {et~al.}(2003){Vaughan}, {Edelson}, {Warwick}, \&
  {Uttley}}]{Vaughan}
{Vaughan}, S., {Edelson}, R., {Warwick}, R.~S., \& {Uttley}, P. 2003, \mnras,
  345, 1271

\bibitem[{{Woo} {et~al.}(2005){Woo}, {Urry}, {van der Marel}, {Lira}, \&
  {Maza}}]{2005ApJ...631..762W}
{Woo}, J.-H., {Urry}, C.~M., {van der Marel}, R.~P., {Lira}, P., \& {Maza}, J.
  2005, \apj, 631, 762

\end{thebibliography}
\bibliographystyle{aa}

\appendix 

\section{Study of the systematics} \label{appSys}

The systematic uncertainties on the EBL measurement with \hess\ brightest blazars are investigated in this appendix. Following \cite{Sinervo}, two sources of systematics arising from ``poorly understood features of the data or analysis technique" (class 2) and two sources of systematics arising ``from uncertainties in the underlying theoretical paradigm" (class 3) are identified. The main class 2 systematic uncertainty is evaluated with Monte Carlo simulated air showers imaged by the detector and passing through the whole chain analysis \citep[see, e.g.,][ and reference therein for a description of the Monte-Carlo simulations]{Crab}. The limited knowledge of the atmospheric conditions is accounted for with a toy model of the detector acceptance and distribution of events. Class 3 systematics are characterized in this study by the choice of template model for the EBL absorption and the selection of the best intrinsic model for each data set. The impact of the latter is evaluated with the data, testing {\it ad hoc} intrinsic models, while the former is compared with a concurrent modelling established by \cite{Dominguez}.

\begin{figure}[h!]
\centering
\includegraphics[width=0.8\linewidth]{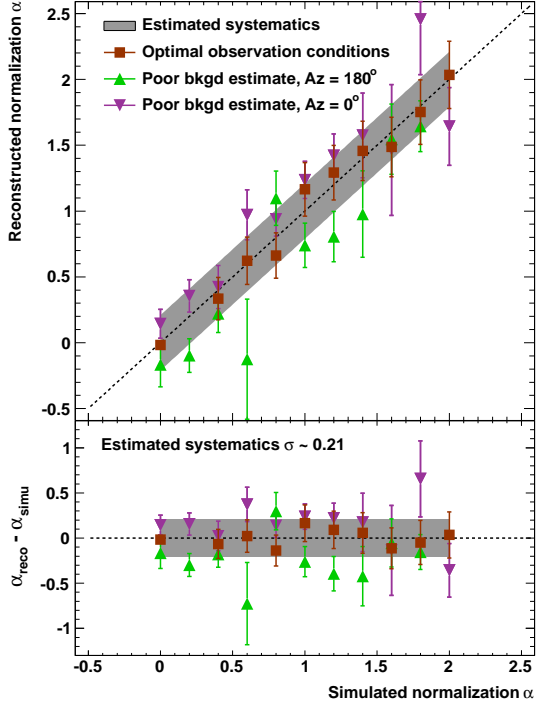}
\caption{Reconstruction of the EBL normalization with Monte Carlo simulated air showers passing through the analysis chain. Three samples of Monte Carlo events are represented: the first one (orange squares) corresponds to the observation conditions of \PKS, the second and third (triangles) correspond to a poor background estimation. These two last sets were used to estimate the systematic uncertainty represented with the grey shaded area. {\it Top panel}: Reconstructed EBL normalization as a function of the simulated normalization. {\it Bottom panel}: Residuals, defined as the difference between the reconstructed and simulated optical depth normalizations.}
\label{fig:SysSoft}
\end{figure}

\subsection{Analysis chain}\label{SysChain}

Monte Carlo data \citep[see][]{Crab} were used to test the analysis chain. Four telescopes triggered events following a PWL of photon index 2 (hardest simulated index) were randomly removed from the simulated data set to create an artificial EBL attenuation. The data set studied was generated for a null azimuth and an off-axis angle of 0.5\dgr. The zenith angle was fixed to 18\dgr, close to the average zenith angle in the H.E.S.S. sky of \PKS, which is the source with the most important excess of $\gamma$-rays in this study (see Sect.~\ref{sample}). The EBL optical depth normalization $\alpha$ was then reconstructed with these samples of events following a spectrum $\phi(E) \propto E^{-2}\exp(-\alpha \times \tau(E,z))$, where $\tau(E,z)$ is the FR08 EBL opacity and $z$ the redshift of the source, fixed here to $z=0.1$ for simplicity.

The background, particularly important for the spectral fit method described in \cite{2001A&A...374..895P}, was fixed to a tenth of the signal - comparable to the value derived for the first data set on \PKS. The reconstructed EBL normalization $\alpha$ is shown in the top panel of Fig.~\ref{fig:SysSoft} as a function of the simulated EBL normalization. The close match with the identity function strongly supports the reliability of the method employed.

The parameter that seems to affect the analysis chain the most is the background estimation, crucial for the mentioned spectral fit method. Imposing a background equivalent to a fiftieth of the signal, two samples of simulated events were studied for a null zenith and respective azimuths of 0\dgr\ and 180\dgr. The azimuth just indexes the data sets, since all azimuth angles are equivalent for a null zenith angle. The corresponding reconstructed EBL normalizations are represented with downward and upward triangles in the top panel of Fig.~\ref{fig:SysSoft}. The associated error bars represent statistical uncertainties, related to the limited size of the Monte Carlo samples (typically $10^4$ events), that must be taken into account when estimating the systematic uncertainty. A first (a priori naive) evaluation of this systematic is the average difference $\alpha_{\rm reco} - \alpha_{\rm simu}$ represented in the bottom panel, which reads 0.17 and 0.20 for each sample. A second evaluation is the maximum variation in the measurement $\Delta$ associated with a Gaussian statistic, yielding one standard deviation systematics $\Delta / \sqrt{12}$ \citep[see, e.g.,][]{Sinervo} of 0.19 and 0.21, respectively. The estimate chosen is similar to the excess variance estimator developed by \cite{Vaughan} for variability. Assuming that the rms difference D between the simulated and reconstructed values is due to both statistical and systematic uncertainties, one would write $\displaystyle D^2={\mathcal V}(\alpha_{\rm reco}-\alpha_{\rm simu})=<\sigma_{\rm stat}^2> + \sigma_{\rm sys}^2$, where ${\mathcal V}$ is the variance estimator. We thus define the systematic uncertainty estimate as:
\begin{equation}
\label{SysEstimate}
\sigma_{\rm sys}=\sqrt{{\mathcal V}(\alpha_{\rm reco}-\alpha_{\rm simu}) - <\sigma_{\rm stat}^2>},
\end{equation}
which reads 0.15 and 0.26 for each sample. The global systematic error using both samples, $\sigma_{\rm sys}=0.21$, is shown in the top and bottom panels of Fig.~\ref{fig:SysSoft}. This systematics estimate is similar to the two mentioned before, though a bit larger, which suggests a possible slight overestimation.

To ensure that a point-to-point systematic effect does not mimic the EBL absorption as a function of energy, a test was performed with a bright galactic source, the Crab Nebula, and yielded deviations to a null EBL normalization well below the systematic uncertainty derived for the analysis chain.

\subsection{Choice of the intrinsic model}\label{IntrinsicModelChoice}

\begin{figure}[h!]
\centering
\includegraphics[width=0.95\linewidth]{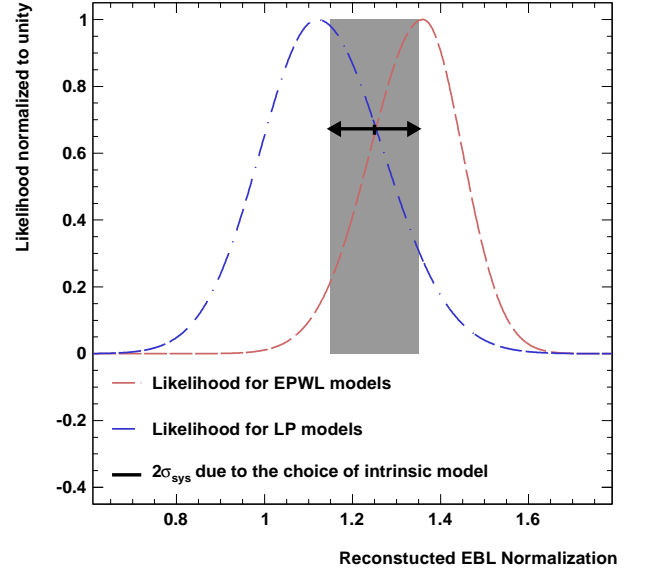}
\caption{Likelihood profiles as a function of the normalized EBL opacity. The profiles were normalized to unity for clarity purposes. The dotted dashed curve is derived fitting log-parabolic intrinsic spectra to the data sets, while the dashed curve is derived by fitting exponential cut-off models. The gap between the two profiles due to the intrinsic spectral modelling is represented by the grey shaded area and the double arrow.}
\label{fig:SysModel}
\end{figure}

The second systematic uncertainty arises from the choice of the model for the intrinsic spectra. This systematic was assessed on the data by comparing the total likelihood profile derived with a LP for each intrinsic spectrum, on one hand, and with an EPWL, on the other. The corresponding likelihoods as a function of the EBL normalization are shown in Fig~\ref{fig:SysModel}, where the maximum was set to unity for clarity. The comparison of third-order models such as ELP and SEPWL would only drown the systematic error in the statistical one. The two profiles were fitted with the procedure described in Sect.~\ref{Results}, yielding $\alpha_{\rm Expcut-off}=1.36^{+0.09}_{-0.12}$ and $\alpha_{\rm LogParabola}=1.12^{+0.15}_{-0.13}$. Using the last systematic estimator described in the Sect.~\ref{SysChain}, the difference between these two values due to the statistics is estimated to 0.14 (variance due to uncertainties), and the deviation caused by the systematics is 0.10.

To ensure the reliability of the measurement, three other selection criteria of the intrinsic model were tested. First, the model with the best $\chi^2$ probability was selected (as in the main method), but the flattest likelihood profile was used in case of ambiguity (e.g. between a LP and an EPWL), yielding a normalization of $1.18 \pm 0.18$, preferred at the $8.9\sigma$ level to a null opacity. A second approach consisted in choosing the simplest model, as long as the next order was not preferred at the $2\sigma$ level (taking the flattest profile in case of ambiguity), yielding a normalization of $1.46 \pm 0.11$, preferred at the $14.3\sigma$ level to a null opacity. These two criteria do not change the intrinsic model for the data sets on \ESn, \ESu, \Mrk\ ({\it 2}), \PKSb\ ({\it 1} and {\it 2}), and \PKS\ ({\it 1}, {\it 6}, and {\it 7}). A final test consisted in imposing the most complex model (an ELP) on the other data sets, yielding a normalization of $1.29 \pm 0.18$, preferred at the $7.9\sigma$ level to a null opacity. The above-mentioned systematic uncertainty accounts for the slight changes induced by the selection method and the significance of the result remains almost unchanged.

It is worth noting that the particular attention paid to the intrinsic curvature of the spectra all along the analysis is not superfluous. The likelihood profile obtained assuming that the spectra are described by PWLs is maximum for $\alpha_{\rm PowerLaw}=2.01\pm0.07$. The value derived with such a basic spectral model is significantly above the nominal normalized EBL opacity because intrinsic curvature of the spectra mimics the absorption effect.

\subsection{Energy scale and choice of the EBL model}\label{EBLmodellingChoice}

The atmosphere is the least understood component of an array of Cherenkov telescopes such as \hess\ and can affect the absorption of the Cherenkov light emitted by the air showers. This absorption leads to a decrease in the number of photoelectrons and thus of the reconstructed energy of the primary $\gamma$-ray. The typical energy shift, of the order of $10\%$ \citep{Bernlohr00}, does not affect the slope of a PWL spectrum, which is energy-scale invariant, but impacts its normalization. Indeed, for an initial spectrum $\phi(E)=\phi_0 (E/E_0)^{-\Gamma}$, an energy shift $\delta$ yields a measured spectrum $\displaystyle \phi_{\rm mes}(E)=\phi_0 [(1+\delta)E/E_0]^{-\Gamma}=\phi_0' (E/E_0)^{-\Gamma} $, where $\phi_0'=(1+\delta)^{-\Gamma} \phi_0$ is the measured spectral normalization. Since the spectral analysis developed in this study relies on the EBL absorption feature which is not an energy-scale invariant spectral model, the atmosphere absorption impact on the measured EBL normalization is investigated.

\begin{figure}[h]
\hspace{-0.2cm}
\includegraphics[width=1.0\linewidth]{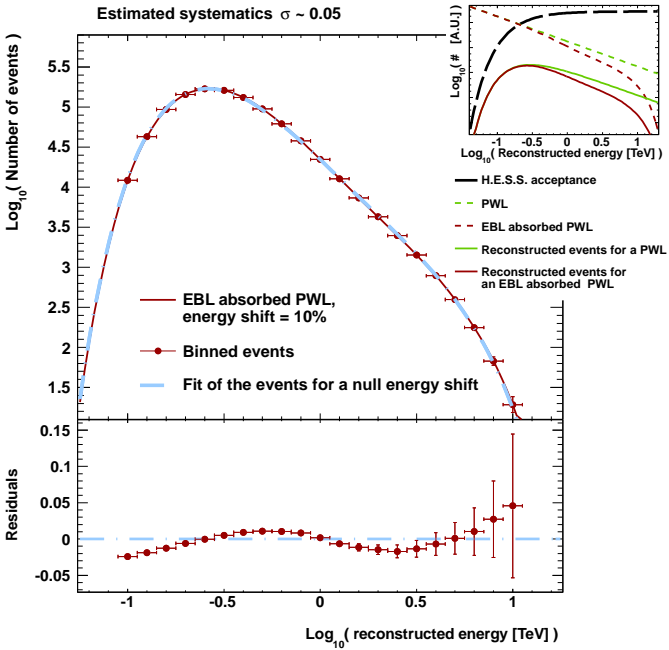}
\caption{Toy-model of the energy distribution of \hess\ events. The inset in the top panel shows the detector acceptance (black line) and the expected distributions of events for a PWL and an EBL absorbed PWL (green and brown lines, respectively). The injected spectra are shifted in energy to model the absorption of Cherenkov light by the atmosphere yielding the distribution of events shown in the top panel with brown filled circles. Fitting this distribution with a non shifted model enables the characterization of the atmospheric impact on the EBL normalization estimated to 0.05 for an energy shift of 10\%. The residuals of the fit are shown in the bottom panel.}
\label{fig:EnergyShift}
\end{figure}

A toy-model of the detector and of the atmosphere effect was developed to account for such a systematic effect. The detector acceptance $\mathcal{A}(E)$ is parametrized as a function that tends to the nominal acceptance value at high energies, as in Eq.~\ref{Eq:AccParam}:
\begin{equation}
\label{Eq:AccParam}
\log_{10} \mathcal{A}(E)=a\times\left[1-b\exp(-c\times\log_{10}E)\right]
\end{equation}
where $\mathcal{A}(E)$ is in m$^2$, the energy $E$ is in TeV, and $a=5.19$, $b=2.32\times10^{-2}$, $c=3.14$ are derived from the fit of the simulated acceptance. The number of events measured in an energy band ${\rm d}E$ is then simply ${\rm d}N / {\rm d}E=\mathcal{A}(E)\times\phi(E)\times T_{obs}$, where the observation duration $T_{obs}$ was fixed to impose a total number of events of $10^6$. Typical event distributions for PWL and EBL absorbed PWL spectra are shown in the inset in Fig.\ref{fig:EnergyShift}. A logarithmic energy binning of $\Delta\log_{10}E=0.1$ is adopted and the uncertainty on the number of events in each energy bin is considered to be Poissonian. To model the effect of the atmosphere on the EBL normalization reconstruction, energy-shifted distributions ${\rm d}N / {\rm d}E=\mathcal{A}(E)\times\phi(E_{\rm shift})\times T_{obs}$ were fitted with an non-shifted model, i.e. $\propto \mathcal{A}(E)\times\phi(E)$, with $E_{\rm shift}=(1+\delta)\times E$ and $\phi(E) \propto E^{-\Gamma} \exp(-\alpha\times\tau(E,z))$. As mentioned above the effect on the index $\Gamma$ is null because of the energy-scale invariance, which is not the case for the specific energy dependence of the EBL opacity. A toy-model distribution that was energy shifted is shown in the top panel of Fig.\ref{fig:EnergyShift} for a redshift $z=0.1$ and an injected EBL normalization $\alpha=1$, corresponding to FR08 EBL modelling. The residuals $\Delta\log_{10}({\rm N_{events}})$ to the fit of a non-shifted model are shown in the bottom panel. 

The reconstructed and injected EBL normalizations differ by less than 0.05 for an energy shift of $10\%$, while the difference can go up to 0.11 for an energy shift of $25\%$. The standard atmospheric conditions required by the data selection motivates the use of the $10\%$ energy shift\footnote{\citet{2010A&A...523A...2M} have even shown that a precision of 5\% on the energy scale can be achieved with atmospheric Cherenkov telescopes.} and thus leads to a systematic error due to Cherenkov light absorption of 0.05. 

This toy model of the detector was also employed to compare independent EBL modellings. To probe a reasonable range of models, the lower and upper bounds on the EBL opacity derived by \cite{Dominguez} were used for the injected spectrum, while FR08 modelling was fitted to the event distribution. The variation in the reconstructed normalization is estimated to be 0.06 for a redshift $z=0.1$. The small amplitude of the systematic effects of the atmosphere and of the EBL modelling choice (respectively 0.05 and 0.06) justifies a posteriori the use of the simple framework described in this section and does not motivate a deeper investigation.

\end{document}